\documentclass[onecolumn,prb]{revtex4}
\usepackage{graphicx}
\usepackage{dcolumn}
\usepackage{bm}
\usepackage{amsmath}

\begin{document}

\preprint{}
\title{Theory of thermal and charge transport in diffusive normal metal /
superconductor junctions }
\author{T. Yokoyama$^1$, Y. Tanaka$^1$, A. A. Golubov$^2$ and Y. Asano$^3$}
\affiliation{$^1$Department of Applied Physics, Nagoya University, Nagoya, 464-8603, Japan%
\\
and CREST, Japan Science and Technology Corporation (JST) Nagoya, 464-8603,
Japan \\
$^2$ Faculty of Science and Technology, University of Twente, 7500 AE,
Enschede, The Netherlands\\
$^3$Department of Applied Physics, Hokkaido University,Sapporo, 060-8628,
Japan}
\date{\today}

\begin{abstract}
Thermal and charge transport in diffusive normal metal(DN) / insulator / $s$%
-, $d$- and $p$-wave superconductor junctions are studied based on
the Usadel equation with the Nazarov's generalized boundary
condition. We  derive a general expression of the
thermal conductance in unconventional superconducting junctions.
Thermal conductance, electric conductance of junctions and their
Lorentz ratio are calculated as a function of resistance in DN, the
Thouless energy, magnetic scattering rate in DN and transparency of
the insulating barrier. We also discuss transport properties for
various orientation angles between the normal to the interface and
the crystal axis of superconductors. It is demonstrated that the
proximity effect does not influence the thermal conductance while
the midgap Andreev resonant states suppress it. Dependencies of the
electrical and thermal conductance on temperature are sensitive to
pairing symmetries and orientation angles. The results imply a
possibility to distinguish one pairing symmetry from another based
on the results of experimental observations.

\end{abstract}

\maketitle



%

%




\section{Introduction}

Thermal and electrical conductances are basic properties of a metal.
The abilities to carry heat and charge currents are related to each
other. At low temperatures thermal conductivity $\kappa $ of a metal
is linear in $T$, i.e., $\kappa \propto T$, while the electrical
conductivity $\sigma $ approaches a constant. As a result, the
Lorentz ratio becomes a universal constant: $L\equiv \kappa /\sigma
T=\pi ^{2}/3e^{2}$. This characteristic feature is called the
Wiedemann-Franz (WF) law~\cite{WF} and has been
observed in various electron systems~\cite%
{Kearney,Castellani,Chester,Smrcka,Ong} which may be described by the Fermi
liquid theory.
A violation of the WF law implies the breakdown of the Fermi liquid
description of the electronic states in a metal. For instance, in the normal
state of high-T$_{c}$ cuprates the violation of the WF law suggests the
non-Fermi liquid description~\cite{Hill}. A validity of the Fermi liquid
picture in cuprates is an open question even now.

The heat and charge currents are also important characteristics of a
superconducting state. Thermal conductivity of a bulk superconductor was
first discussed theoretically by Bardeen et al.~\cite{Badeen}. Large amount
of work was done on thermal and electric transport in contacts between
normal and superconducting phases (N/S junctions). In the pioneering work by
Andreev \cite{Andreev} a new type of quasiparticle scattering at the N/S
interface was discovered, the so-called Andreev reflection (AR), which
crucially influences quasiparticle transport across the interface at subgap
energies. The AR causes the exponential decay of a thermal
conductance across the N/S interface with decreasing temperature \cite%
{Andreev}, while it facilitates the transfer of an electric charge\ \cite%
{BTK}. As shown by Blonder, Tinkham and Klapwijk (BTK)\ \cite{BTK}, the
AR leads to the doubling of zero bias conductance across
transparent N/S interface at low $T$. More recently, Bardas and Averin\cite%
{Bardas} and Devyatov \textit{et al.}~\cite{Devyatov} calculated heat
current by generalizing the BTK model~\cite{BTK} for the electric transport
in N/S junctions.
The applicability of these theories is, however, limited to junctions in the
clean limit.

In most of practical N/S junctions normal metals are in the diffusive
 regime. Therefore the effect of impurity scattering on the
transport properties has received a lot of attention. In diffusive
normal metal / superconductor (DN/S) junctions, the diffusive motion
of quasiparticles at a mesoscopic length scale around the interface
strongly modifies the
transport across a junction interface because of interference effects~\cite%
{Hekking}. For example, the appearance of the zero bias conductance peak
(ZBCP) in DN/S junctions is a direct consequence of the interference effect
of a quasiparticle in DN~\cite%
{Giazotto,Klapwijk,Kastalsky,Nguyen,Wees,Nitta,Bakker,Xiong,Magnee,Kutch,Poirier}%
. 
To study such interference effects, the quasiclassical Green's function
theory~\cite{Eilenberger,Eliashberg,Larkin} has been widely used because of
its convenience and broad applicability. Based on this formalism, theory of
charge transport in DN/S junctions was formulated by Volkov, Zaitsev and
Klapwijk (VZK) \cite{Volkov}
. By applying the VZK theory, a number of authors investigated
theoretically charge transport in various proximity structures~\cite
{Lambert1,Nazarov1,Yip,Stoof,Reentrance,Golubov,Takayanagi,Seviour,Belzig,Yip2,Yoko,Volkov}.
This work was based on the boundary conditions for the Keldysh-Nambu
Green's function at the DN/S interface derived by Kupriyanov and
Lukichev (KL)~\cite{KL} from Zaitsev's boundary
condition~\cite{Zaitsev} in the isotropic limit. The KL boundary
conditions were recently extended by Nazarov within the circuit
theory~\cite{Nazarov2} and applied by Tanaka \textit{et 
al.}~\cite{TGK} to the study of
charge transport in N/S junctions with arbitrary interface
transparency in order to investigate more complex structures. The  Nazarov's boundary conditions coincide with the
KL boundary conditions when transmission coefficients are
sufficiently low, while the BTK theory \cite{BTK} is reproduced in
the ballistic regime. It was shown in \cite{TGK} that a zero bias
conductance peak (ZBCP) in low transparent N/S junctions transforms
to a zero bias conductance dip (ZBCD) with increasing the interface
transparency.

Heat transport within the quasiclassical approach was studied by several
authors~\cite{Raimondi,Beyer,Graf,Graf2,Bezuglyi}. In particular, Graf
\textit{et al.}~\cite{Graf} calculated thermal conductivity for various
unconventional superconductors. They showed that thermal conductivity of a
clean two-dimensional $d_{x^{2}-y^{2}}$-wave superconductor is proportional
to $T$ in the Born limit over a broad temperature range and proportional to $%
T^{3}$ in the unitary limit above some crossover temperature $T^{\ast }\sim
\gamma $, where $\gamma $ is the bandwidth of quasiparticle states bound to
impurities. These results also suggest that thermal conductivity is
sensitive to a pairing symmetry of an unconventional superconductor~\cite%
{Graf2}.
On the other hand, heat transport in N/S junctions has not received much
attention so far, partially due to the lack of experimental data. Recently,
sufficient progress was achieved which made it possible to study heat
transport in unconventional superconductors \cite%
{Movshovich,Movshovich2,May,Izawa,Izawa2,Izawa3,Izawa4,Suzuki} and this
stimulated theoretical study of these phenomena.

\begin{figure}[tbh]
\begin{center}
\scalebox{0.4}{
\includegraphics[width=22.0cm,clip]{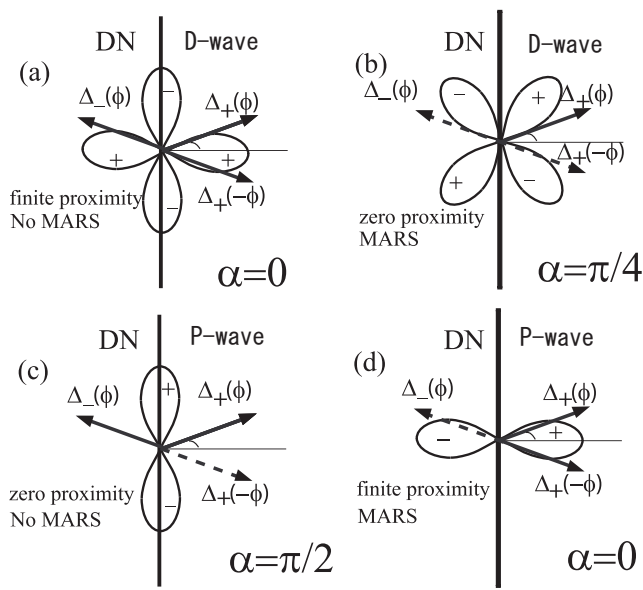}
}
\end{center}
\caption{ Schematic illustration of trajectories of incoming and outgoing
quasiparticles at a DN/S interface with pair potential $\Delta _{\pm }(%
\protect\phi )$. The pair potentials $\Delta _{\pm }$ are given by $\Delta
_{\pm }=\Delta (T)\cos [2(\protect\phi \mp \protect\alpha )]$ for $d$-wave
superconductors (D-wave) where $\protect\alpha $ denotes the angle between
the normal to the interface and the crystal axis of $d$-wave
superconductors. For $p$-wave superconductors (P-wave) the pair potentials $%
\Delta _{\pm }$ are given by $\Delta _{\pm }=\pm \Delta (T)\cos [(\protect%
\phi \mp \protect\alpha )]$, where $\protect\alpha $ denotes the angle
between the normal to the interface and the lobe direction of the $p$-wave
pair potential. In the above, $\protect\phi $ denotes the injection angle of
the quasiparticle measured from the $x$-axis and $\Delta (T)$ is the maximum
amplitude of the pair potential at a temperature $T$.}
\label{f0}
\end{figure}

The formation of the midgap Andreev resonant state (MARS) drastically
affects low energy transport in unconventional superconducting junctions~%
\cite{Buch,TK95,Kashi00,Yamashiro,Maeno}. It is now widely accepted
that MARS is also responsible for ZBCP in N/S junctions of unconventional
superconductors. To discuss interplay between the proximity effect and MARS,
a new circuit theory for unconventional superconductors was proposed in~\cite%
{Nazarov2003,TNGK,p-wave}. In DN/$d$-wave superconductor junctions, MARS
interfere destructively with the proximity effect in DN~\cite%
{Nazarov2003,TNGK}. On the other hand, in DN/$p$-wave superconductor
junctions the MARS and the proximity effect coexist~\cite{p-wave}. As a
result, the conductance spectrum has a giant ZBCP and the local density of
states in DN has a zero energy peak. We summarize relations between the
proximity effect and the MARS in Fig.~\ref{f0}.

The junctions in Fig.~\ref{f0} are classified into four groups: (a)
presence of the proximity effect, (b) presence of MARS, (c) absence
of both MARS and the proximity effect, and (d) presence of both of
them. Thermal and charge transport in N/S junctions can be described
in terms of MARS and the proximity effect.

The purpose of the present paper is to study the tunneling conductance, the
thermal conductance and their Lorentz ratio in diffusive normal metal /
insulator / $s$-, $d$- and $p$-wave superconductor junctions as a function
of transparencies of insulating barriers at the interface, resistance $R_{d}$
in DN, the magnetic scattering rate in DN, the Thouless energy $E_{Th}$ in
DN, and orientation angles between the normal to the interface and the
crystal axis of superconductors.

The organization of this paper is as follows. In section II, we will provide
the detailed derivation of the expression for the normalized thermal
conductance. In section III, the results of calculations are presented for
various types of junctions. They are applied to discriminate various pairing
states. In section IV, the summary of the obtained results is given.
In the present paper, we use the units with $k_{B}=\hbar =1$.

\section{Formulation}

In this section, we explain the model and the formalism. We consider a
junction consisting of normal and superconducting reservoirs connected by a
quasi-one-dimensional diffusive conductor with a length $L$ much larger than
the mean free path. The interface between the DN and the S
has a resistance $R_{b}$ while the DN/N interface has zero resistance. The
positions of the DN/N interface and the DN/S interface are denoted as $x=0$
and $x=L$, respectively. According to the circuit theory, the interface
between DN and S is subdivided into two isotropization zones in DN and S,
two ballistic zones and a scattering zone. The sizes of the ballistic and
scattering zones in the current direction are much shorter than the
coherence length.
The scattering zone is modeled as an infinitely narrow insulating barrier
described by the delta function $U(x)=H\delta (x-L)$. The transparency of
the interface $T_{n}$ is given by $T_{n}=4\cos ^{2}\phi /(4\cos ^{2}\phi
+Z^{2})$, where $Z=2H/v_{F}$ is a dimensionless parameter, $\phi $ is an
injection angle measured from the interface normal and $v_{F}$ is the Fermi
velocity. The interface resistance $R_{b}$ is given by
\begin{equation*}
R_{b}=R_{0}\frac{2}{\int_{-\pi /2}^{\pi /2}d\phi T_{n}\cos \phi },
\end{equation*}%
where $R_{0}$ is the Sharvin resistance $R_{0}^{-1}=e^{2}k_{F}^{2}S_{c}/4\pi
^{2}$, $k_{F}$ is the Fermi wave length, and $S_{c}$ is the constriction
area. 

We apply the quasiclassical Keldysh formalism in the following calculation
of the tunneling and thermal conductance. The 4 $\times $ 4 Green's
functions in DN and S are denoted by $\check{G}_{1}(x)$ and $\check{G}%
_{2}(x) $, respectively. The spatial dependence of $\check{G}_{1}(x)$ in DN
is determined by the static Usadel equation~\cite{Usadel},
\begin{equation}
D\frac{\partial }{\partial x}[\check{G}_{1}(x)\frac{\partial \check{G}_{1}(x)%
}{\partial x}]+i[\check{H}+i\check{\Sigma}_{spin},\check{G}_{1}(x)]=0
\end{equation}%
with the diffusion constant $D$ in DN, where $\check{H}$ is given by
\begin{equation*}
\check{H}=\left(
\begin{array}{cc}
\hat{H}_{0} & 0 \\
0 & \hat{H}_{0}%
\end{array}%
\right) ,
\end{equation*}%
with $\hat{H}_{0}=\epsilon \hat{\tau}_{3}$, $\check{\Sigma}_{spin}=\frac{%
\gamma }{2}\hat{\tau}_{3}\check{G}_{1}(x)\hat{\tau}_{3}$ is the self-energy
for magnetic impurity scattering with the scattering rate $\gamma $ and $%
\epsilon $ is the quasiparticle energy. The directions of magnetic moments
of impurities are random. The self-energy is given by averaging with respect
to directions of magnetic moments. Thus in our calculation $\check{G}_{1}(x)$
is unit matrix in spin space.

The electric current is expressed using $\check{G}_{1}(x)$ as
\begin{equation}
I_{el}=\frac{-L}{4eR_{d}}\int_{0}^{\infty }d\epsilon \mathrm{Tr}[\hat{\tau}
_{3}(\check{G}_{1}(x)\frac{\partial \check{G}_{1}(x)}{\partial x})^{K}],
\label{Electric}
\end{equation}%
where $(\check{G_{1}}(x)\frac{\partial \check{G_{1}}(x)}{\partial x})^{K}$
denotes the Keldysh component of $(\check{G_{1}}(x)\frac{\partial \check{%
G_{1}}(x)}{\partial x})$.

The thermal current is also expressed as
\begin{equation}
I_{th}=\frac{L}{4e^{2}R_{d}}\int_{0}^{\infty }d\epsilon \epsilon \mathrm{Tr}%
[(\check{G}_{1}(x)\frac{\partial \check{G}_{1}(x)}{\partial x})^{K}].
\label{Thermal}
\end{equation}%
%
%
%
%
%
It is convenient to use the standard $\theta $-parametrization when function
$\hat{R}_{1}(x)$ is expressed as
\begin{equation}
\hat{R}_{1}(x)=\hat{\tau}_{1}\sin \theta (x)\cos \psi +\hat{\tau}_{2}\sin
\theta (x)\sin \psi +\hat{\tau}_{3}\cos \theta (x)
\end{equation}%
where $\cos \psi =0$ for singlet superconductors and $\sin \psi =0$ for
triplet superconductors(see Ref.~\cite{p-wave}). The parameter $\theta (x)$
is a measure of the proximity effect in DN.

Functions $\hat{A}_{1}(x)$ and $\hat{K}_{1}(x)$ are expressed as $\hat{ A}
_{1}(x)=-\hat{\tau}_{3}\hat{R}_{1}^{\dag}(x)\hat{\tau}_{3}$ and $\hat{K}%
_{1}(x)=\hat{R}_{1}(x)\hat{f } _{1}(x)-\hat{f}_{1}(x)\hat{A}_{1}(x)$ with
the distribution function $\hat{ f} _{1}(x)$ which is given by $\hat{f}%
_{1}(x)=f_{l}(x)+\hat{\tau}_{3}f_{t}(x)$. From the retarded or advanced
component of the Usadel equation, the spatial dependence of $\theta (x)$ is
determined by the following equation
\begin{equation}
D\frac{\partial ^{2}}{\partial x^{2}}\theta (x)+2i(\epsilon + i\gamma \cos
[\theta (x)])\sin [\theta (x)]=0,  \label{Usa1}
\end{equation}
while from the Keldysh component we obtain
\begin{equation}
D\frac{\partial }{\partial x}[\frac{\partial f_{l}(x)}{\partial x} \mathrm{%
cos^{2}}\mathrm{Re}[\theta(x)]]=0.  \label{Usa2}
\end{equation}%
The average over injection angles of quasiparticles at the interface is
defined as
\begin{equation*}
<B(\phi)> = \int_{-\pi/2}^{\pi/2} d\phi \cos\phi B(\phi)
/\int_{-\pi/2}^{\pi/2} d\phi T(\phi)\cos\phi
\end{equation*}
with $T(\phi)=T_{n}$.

\subsection{$s$-wave case}

The Keldysh component $\hat{K}_{2}(x)$ is given by $\hat{K}_{2}(x)=\hat{R}%
_{2}(x)\hat{f}_{2}(x)-\hat{f}_{2}(x)\hat{A}_{2}(x)$ with the retarded
component $\hat{R}_{2}(x)$, the advanced component $\hat{A}_{2}(x)$ and
distribution function $\hat{f}_{2}(x)$. Here $\hat{R}_{2}(x)=g\hat{\tau}%
_{3}+f\hat{\tau}_{2}$ with $g=\epsilon /\sqrt{\epsilon ^{2}-\Delta ^{2}(T)}$
and $f=\Delta (T)/\sqrt{\Delta ^{2}(T)-\epsilon ^{2}}$, $\hat{A}_{2}(x)=-%
\hat{\tau}_{3}\hat{R}_{2}^{\dagger }(x)\hat{\tau}_{3}$ and $\hat{f}%
_{2}(x)=f_{S}=\mathrm{{tanh}[\epsilon /(2T)]}$ in thermal equilibrium with
temperature $T$. 
The boundary condition for $\check{G}_{1}(x)$ at the DN/S interface is given
by\cite{Nazarov2},
\begin{equation}
\frac{L}{R_{d}}(\check{G}_{1}\frac{\partial \check{G}_{1}}{\partial x}%
)_{\mid x=L_{-}}=R_{b}^{-1}<B>,  \label{Nazarov}
\end{equation}%
\begin{equation*}
B=\frac{2T_{n}[\check{G}_{1}(L_{-}),\check{G}_{2}(L_{+})]}{4+T_{n}([\check{G}%
_{1}(L_{-}),\check{G}_{2}(L_{+})]_{+}-2)}.
\end{equation*}%
For the electrical conductance, we obtain the following result at zero
voltage~\cite{TGK}
\begin{equation}
\sigma _{S}(T)=\frac{1}{2T}\int_{0}^{\infty }\frac{d\epsilon }{\cosh ^{2}[%
\frac{\epsilon }{2T}](\frac{R_{b}}{<I_{b0}>}+\frac{R_{d}}{L}\int_{0}^{L}%
\frac{dx}{\cosh ^{2}\theta _{im}(x)})} \label{cond}
\end{equation}%
with
\begin{equation*}
I_{b0}=\frac{T_{n}^{2}\Lambda _{1}+2T_{n}(2-T_{n})\Lambda _{2}}{2\mid
(2-T_{n})+T_{n}[g\cos \theta _{L}+f\sin \theta _{L}]\mid ^{2}},
\end{equation*}%
\begin{equation*}
\Lambda _{1}=(1+\mid \cos \theta _{L}\mid ^{2}+\mid \sin \theta _{L}\mid
^{2})(\mid g\mid ^{2}+\mid f\mid ^{2}+1)
\end{equation*}%
\begin{equation}
+4\mathrm{Imag}[fg^{\ast }]\mathrm{Imag}[\cos \theta _{L}\sin \theta
_{L}^{\ast }],
\end{equation}

\begin{equation}
\Lambda _{2}=\mathrm{{Real}\{g(\cos \theta _{L}+\cos \theta _{L}^{\ast
})+f(\sin \theta _{L}+\sin \theta _{L}^{\ast })\}},
\end{equation}%
where $\theta _{im}(x)$ and $\theta _{L}$ denote the imaginary parts of $%
\theta (x)$ and $\theta (L_{-})$ respectively.

Next we calculate the thermal conductance.
Since DN is attached to the normal electrode at $x=0$, $\theta (0)$=0 and $%
f_{l}(0)=f_{l0}$ with $f_{l0}=\tanh [\epsilon /(2(T+\Delta T))]$ in thermal
equilibrium at a temperature $T+\Delta T$.

The retarded part of Eq.~(\ref{Nazarov}) is the same as in Ref.~%
\onlinecite{TGK}. The Keldysh part of Eq.~(\ref{Nazarov}) reads
\begin{equation}
\frac{L}{R_{d}}(\frac{\partial f_{l}}{\partial x})\mathrm{{\cos ^{2}}}%
\mathrm{Real}[\theta (x)]\mid _{x=L_{-}}=-\frac{<I_{b1}>(f_{l}(L_{-})-f_{S})%
}{R_{b}},  \label{b2}
\end{equation}%
with
\begin{equation*}
I_{b1}=\frac{T_{n}^{2}\Lambda _{1}^{^{\prime }}+T_{n}(2-T_{n})\Lambda
_{2}^{^{\prime }}}{2\mid (2-T_{n})+T_{n}[g\cos \theta _{L}+f\sin \theta
_{L}]\mid ^{2}},
\end{equation*}%
\begin{equation*}
\Lambda _{1}^{^{\prime }}=(1+\mid \cos \theta _{L}\mid ^{2}-\mid \sin \theta
_{L}\mid ^{2})(\mid g\mid ^{2}-\mid f\mid ^{2}+1)
\end{equation*}%
\begin{equation}
+4\mathrm{Real}[fg^{\ast }]\mathrm{Real}[\cos \theta _{L}\sin \theta
_{L}^{\ast }],
\end{equation}

\begin{equation}
\Lambda _{2}^{^{\prime }}=4\mathrm{Real}[g]\mathrm{Real}[\cos \theta _{L}]-4%
\mathrm{Imag}[f]\mathrm{Imag}[\sin \theta _{L}].
\end{equation}%
%
%
%
%
%
Substituting the results into Eq.~(\ref{Thermal}), we arrive at the final
expression for the thermal current
\begin{equation}
I_{th}=\frac{1}{{e^{2}}}\int_{0}^{\infty }{\frac{{\epsilon \left( {%
f_{l0}-f_{S}}\right) d\epsilon }}{{\frac{{R_{b}}}{{\left\langle {I_{b1}}%
\right\rangle }}+\frac{{R_{d}}}{L}\int\limits_{0}^{L}{\frac{{dx}}{{\cos ^{2}{%
\mathop{\rm Re}\nolimits}\theta \left( x\right) }}}}}}
\end{equation}%
and the thermal conductance%
\begin{equation}
\kappa =\mathop {\lim }\limits_{\Delta T\rightarrow 0}\frac{{I_{th}}}{{%
\Delta T}}=\frac{1}{{2e^{2}T^{2}}}\int_{0}^{\infty }{\frac{{\epsilon 
^{2}d\epsilon }}{{\cosh ^{2}\left( {\frac{\epsilon }{{2T}}}\right)
\left( {\frac{{R_{b}}}{{\left\langle {I_{b1}}\right\rangle }}+\frac{{R_{d}}}{%
L}\int\limits_{0}^{L}{\frac{{dx}}{{\cos ^{2}{\mathop{\rm Re}\nolimits}\theta
\left( x\right) }}}}\right) }}}.  \label{thermal}
\end{equation}

\subsection{$d$-wave case}

In the following $\pm $ stands for the direction of motion along the $x$
axis. We denote Keldysh-Nambu Green's function $\check{G}_{2\pm }$ as follows,
\begin{equation}
\check{G}_{2\pm }=\left(
\begin{array}{cc}
\hat{R}_{2\pm } & \hat{K}_{2\pm } \\
0 & \hat{A}_{2\pm }%
\end{array}%
\right) ,
\end{equation}%
where the Keldysh component $\hat{K}_{2\pm }$ is given by $\hat{K}_{2\pm }=%
\hat{R}_{2\pm }\hat{f}_{2}(x)-\hat{f}_{2}(x)\hat{A}_{2\pm }$ with the
retarded component $\hat{R}_{2\pm }$, the advanced component $\hat{A}_{2\pm
} $ and the distribution function $\hat{f}_{2}(x)$. Here $\hat{R}_{2\pm
}=g_{\pm }\hat{\tau}_{3}+f_{\pm }\hat{\tau}_{2}$ with $g_{\pm }=\epsilon /%
\sqrt{\epsilon ^{2}-\Delta _{\pm }^{2}(T)}$, $f_{\pm }=\Delta _{\pm }(T)/%
\sqrt{\Delta _{\pm }^{2}(T)-\epsilon ^{2}}$ and $\hat{A}_{2\pm }=-\hat{%
\tau}_{3}\hat{R}_{2\pm }^{\dagger }\hat{\tau}_{3}$. The function $\hat{f}%
_{2}(x)$ is given in the previous subsection. For the electrical
conductance, following Ref.\cite{TNGK}, we obtain the conductance, $\sigma_{S}(T)$, at zero voltage by replacing $I_{b2}$ into $I_{b0}$ in Eq.~(\ref{cond})
where
\begin{equation*}
I_{b2}=\frac{T_{n}}{2}\frac{C_{0}}{\mid
(2-T_{n})(1+g_{+}g_{-}+f_{+}f_{-})+T_{n}[\cos \theta _{L}(g_{+}+g_{-})+\sin
\theta _{L}(f_{+}+f_{-})]\mid ^{2}}
\end{equation*}%
\begin{equation*}
C_{0}=T_{n}(1+\mid \cos \theta _{L}\mid ^{2}+\mid \sin \theta _{L}\mid
^{2})[\mid g_{+}+g_{-}\mid ^{2}+\mid f_{+}+f_{-}\mid ^{2}+\mid
1+f_{+}f_{-}+g_{+}g_{-}\mid ^{2}+\mid f_{+}g_{-}-g_{+}f_{-}\mid ^{2}]
\end{equation*}%
\begin{equation*}
+2(2-T_{n})\mathrm{Real}\{(1+g_{+}^{\ast }g_{-}^{\ast }+f_{+}^{\ast
}f_{-}^{\ast })[(\cos \theta _{L}+\cos \theta _{L}^{\ast
})(g_{+}+g_{-})+(\sin \theta _{L}+\sin \theta _{L}^{\ast })(f_{+}+f_{-})]\}
\end{equation*}%
\begin{equation*}
+4T_{n}\mathrm{Imag}(\cos \theta _{L}\sin \theta _{L}^{\ast })\mathrm{Imag}%
[(f_{+}+f_{-})(g_{+}^{\ast }+g_{-}^{\ast })].
\end{equation*}

Let us calculate the thermal conductance. The boundary condition for $\check{%
G}_{1}(x)$ at the DN/S interface reads\cite{Nazarov2}
\begin{equation}
\frac{L}{R_{d}}(\check{G}_{1}\frac{\partial \check{G}_{1}}{\partial x}%
)_{\mid x=L_{-}}=R_{b}^{-1}<\check{I}_{n}>.  \label{Nazarov3}
\end{equation}%
Here $\check{I}_{n}$ is given in the Appendix.
The retarded part of Eq.~(\ref{Nazarov3}) is the same as that in Ref.~%
\onlinecite{TNGK}. The Keldysh part of Eq.~(\ref{Nazarov3}) reads (see the
Appendix)
\begin{equation}
\frac{L}{R_{d}} (\frac{\partial f_{l}}{ \partial x}) \mathrm{{\cos^{2}}}
\mathrm{Real}[\theta(x)] \mid_{x=L_{-}} =-\frac{<I_{b3}> (f_{l}(L_{-})-f_S)}{%
R_{b}},  \label{b3}
\end{equation}
with
\begin{equation*}
I_{b3}=\frac{T_{n}}{2}\frac{C_{0}^{\prime}}{\mid
(2-T_{n})(1+g_{+}g_{-}+f_{+}f_{-})+T_{n}[\cos \theta _{L}(g_{+}+g_{-})+\sin
\theta _{L}(f_{+}+f_{-})]\mid ^{2}}.
\end{equation*}
For isotropic limit, where $f_{+}=f_{-}$ and $g_{+}=g_{-}$ are satisfied, we
obtain $I_{b3}=I_{b1}$.

Applying a similar procedure as in the $s$-wave case, we obtain the thermal
conductance  by replacing $I_{b3}$ into $I_{b1}$ in Eq.~(\ref{thermal})

For the ballistic limit, where $\theta _{L}=R_{d}=0$, we can reproduce the
formula for the thermal conductance for an ideal interface ($T_{n}=1$, i.e.,
$Z=0$) found in the previous work~\cite{Devyatov}:
\begin{equation*}
\kappa =\frac{1}{{2e^{2}T^{2}}}\int_{0}^{\infty }{\frac{\epsilon 
^{2}<I_{b3}>d\epsilon }{R_{b}\cosh ^{2}\frac{\epsilon }{2T}}}
\end{equation*}%
where
\begin{equation}
I_{b3}=1-\frac{\mid \Gamma _{+}\mid ^{2}+\mid \Gamma _{-}\mid ^{2}}{2}
\end{equation}%
with
\begin{equation*}
\Gamma _{+}=\frac{\Delta _{+}}{\epsilon +\sqrt{\epsilon ^{2}-\Delta
_{+}^{2}}},\ \Gamma _{-}=\frac{\Delta _{-}}{\epsilon +\sqrt{\epsilon 
^{2}-\Delta _{-}^{2}}}.
\end{equation*}


\subsection{$p$-wave case}

Here, we restrict our attention to $p$-wave superconductors with $S_{z}=0$,
where $S_{z}$ denotes the $z$ component of the total spin of a Cooper pair.
In this case the derivation is similar to that of the $d$-wave case. In the
following, we will use the same notations as in the $d$-wave case. We can
choose $\hat{R}_{1}=\cos \theta (x)\hat{\tau}_{3}+\sin \theta (x)\hat{\tau}%
_{1}$ to satisfy the boundary condition at the interface. $2\times 2$ matrix
$C_{i}$ $(i=1-6)$ is expressed in terms of linear combination of $\hat{1}$, $%
\hat{\tau}_{1}$, $\hat{\tau}_{2}$ and $\hat{\tau}_{3}$, and $\hat{B}%
_{R}=b_{1}\hat{\tau}_{1}+b_{2}\hat{\tau}_{2}+b_{3}\hat{\tau}_{3}$ with
\begin{equation*}
b_{1}=\frac{-T_{1n}[T_{1n}\sin \theta _{L}+i(f_{+}g_{-}-f_{-}g_{+})]}{%
(1+T_{1n}^{2})(1+g_{+}g_{-}+f_{+}f_{-})+2T_{1n}[\cos \theta
_{L}(g_{+}+g_{-})+i\sin \theta _{L}(f_{+}g_{-}-f_{-}g_{+})]},
\end{equation*}%
\begin{equation*}
b_{2}=\frac{-T_{1n}[T_{1n}\sin \theta
_{L}(1+g_{+}g_{-}+f_{+}f_{-})+i(f_{+}g_{-}-f_{-}g_{+})]}{%
(1+T_{1n}^{2})(1+g_{+}g_{-}+f_{+}f_{-})+2T_{1n}[\cos \theta
_{L}(g_{+}+g_{-})+i\sin \theta _{L}(f_{+}g_{-}-f_{-}g_{+})]},
\end{equation*}%
\begin{equation}
b_{3}=\frac{-T_{1n}[T_{1n}\cos \theta
_{L}(1+g_{+}g_{-}+f_{+}f_{-})+g_{+}+g_{-}]}{%
(1+T_{1n}^{2})(1+g_{+}g_{-}+f_{+}f_{-})+2T_{1n}[\cos \theta
_{L}(g_{+}+g_{-})+i\sin \theta _{L}(f_{+}g_{-}-f_{-}g_{+})]}.
\end{equation}

For the electrical conductance, following Ref.~\onlinecite{p-wave},  we obtain the conductance at zero voltage by replacing $I_{b4}$ into $I_{b0}$ in Eq.~(\ref{cond})
with
\begin{equation*}
I_{b4}=\frac{T_{n}}{2}\frac{C_{0}^{\prime \prime }}{\mid
(2-T_{n})(1+g_{+}g_{-}+f_{+}f_{-})+T_{n}[\cos \theta _{L}(g_{+}+g_{-})+i\sin
\theta _{L}(f_{+}g_{-}-g_{+}f_{-})]\mid ^{2}}
\end{equation*}%
\begin{equation*}
C_{0}^{\prime \prime }=T_{n}(1+\mid \cos \theta _{L}\mid ^{2}+\mid \sin
\theta _{L}\mid ^{2})[\mid g_{+}+g_{-}\mid ^{2}+\mid f_{+}+f_{-}\mid
^{2}+\mid 1+f_{+}f_{-}+g_{+}g_{-}\mid ^{2}+\mid f_{+}g_{-}-g_{+}f_{-}\mid
^{2}]
\end{equation*}%
\begin{equation*}
+2(2-T_{n})\mathrm{Real}\{(1+g_{+}^{\ast }g_{-}^{\ast }+f_{+}^{\ast
}f_{-}^{\ast })[(\cos \theta _{L}+\cos \theta _{L}^{\ast
})(g_{+}+g_{-})+i(\sin \theta _{L}+\sin \theta _{L}^{\ast
})(f_{+}g_{-}-g_{+}f_{-})]\}
\end{equation*}%
\begin{equation*}
+4T_{n}\mathrm{Imag}(\cos \theta _{L}\sin \theta _{L}^{\ast })\mathrm{Imag}%
[i(f_{+}g_{-}-g_{+}f_{-})(g_{+}^{\ast }+g_{-}^{\ast })].
\end{equation*}

Next we proceed with the discussion of the thermal conductance. The retarded
part of Eq.~(\ref{Nazarov3}) is the same as that in Ref.~\onlinecite{p-wave}%
. We have to calculate the Keldysh part of Eq.~(\ref{Nazarov3}) as in the $d$%
-wave case. Following equations are also satisfied:

\begin{equation}
\hat{D}_{R}^{-1}(T_{1n}-T_{1n}\hat{R}_{m}^{-1}+T_{1n}^{2}\hat{R}_{1}\hat{R}%
_{m}^{-1}\hat{R}_{p})=\hat{B}_{R},
\end{equation}

\begin{equation}
\hat{B}_{R} (1 + \hat{R}_{m}^{-1} + T_{1n}\hat{R}_{1}\hat{R}_{p}\hat{R}%
_{m}^{-1}) =T_{1n}\hat{R}_{m}^{-1}\hat{R}_{p}.
\end{equation}

Applying similar procedure as in the $d$-wave case, we obtain the following
expression for the thermal conductance by replacing $I_{b5}$ into $I_{b1}$ in Eq.~(\ref{thermal})
where
\begin{equation*}
I_{b5}=\frac{T_{n}}{2}\frac{C_{0}^{\prime \prime \prime }}{\mid
(2-T_{n})(1+g_{+}g_{-}+f_{+}f_{-})+T_{n}[\cos \theta _{L}(g_{+}+g_{-})+i\sin
\theta _{L}(f_{+}g_{-}-f_{-}g_{+})]\mid ^{2}}
\end{equation*}%
\begin{equation*}
C_{0}^{\prime \prime \prime }=T_{n}(1+\mid \cos \theta _{L}\mid ^{2}-\mid
\sin \theta _{L}\mid ^{2})(\mid g_{+}+g_{-}\mid ^{2}-\mid f_{+}+f_{-}\mid
^{2}+\mid 1+f_{+}f_{-}+g_{+}g_{-}\mid ^{2}-\mid f_{+}g_{-}-g_{+}f_{-}\mid
^{2})
\end{equation*}%
\begin{equation*}
+4(2-T_{n})[\mathrm{Real}[{(1+g_{+}^{\ast }g_{-}^{\ast }+f_{+}^{\ast
}f_{-}^{\ast })(g_{+}+g_{-})}]\mathrm{Real}(\cos \theta _{L})-\mathrm{Imag}%
(\sin \theta _{L})\mathrm{Imag}[{(1+g_{+}^{\ast }g_{-}^{\ast }+f_{+}^{\ast
}f_{-}^{\ast })(i(f_{+}g_{-}-f_{-}g_{+}))}]]
\end{equation*}%
\begin{equation*}
+4T_{n}\mathrm{Real}(\cos \theta _{L}\sin \theta _{L}^{\ast })\mathrm{Real}%
[i(f_{+}g_{-}-f_{-}g_{+})(g_{+}^{\ast }+g_{-}^{\ast })].
\end{equation*}

Taking into account all the above results, we can find simpler
expressions for $<I_{bi}>(i=2-5)$ for any triplet superconductors (TS) and
unconventional singlet superconductors (USS):
\begin{equation}
<I_{bi}>=<\frac{T_{n}}{2}\frac{D_{0}}{\mid (2-T_{n})+T_{n}(\cos \theta
_{L}g_{S}+\sin \theta _{L}f_{S})\mid ^{2}}>  \label{ib0}
\end{equation}%
with
\begin{equation*}
D_{0}=T_{n}(1+\mid \cos \theta _{L}\mid ^{2}+\mid \sin \theta _{L}\mid
^{2})[\mid g_{S}\mid ^{2}+\mid f_{S}\mid ^{2}+1+\mid \bar{f}_{S}\mid ^{2}]
\end{equation*}%
\begin{equation*}
+4(2-T_{n})[\mathrm{Real}(g_{S})\mathrm{Real}(\cos \theta _{L})+\mathrm{Real}%
(f_{S})\mathrm{Real}(\sin \theta _{L})]
\end{equation*}%
\begin{equation}
+4T_{n}[\mathrm{Imag}(\cos \theta _{L}\sin \theta _{L}^{\ast })\mathrm{Imag}%
(f_{S}g_{S}^{\ast })]  \label{cond1}
\end{equation}%
for $i=2,4$ , or
\begin{equation*}
D_{0}=T_{n}(1+\mid \cos \theta _{L}\mid ^{2}-\mid \sin \theta _{L}\mid
^{2})[\mid g_{S}\mid ^{2}-\mid f_{S}\mid ^{2}+1-\mid \bar{f}_{S}\mid ^{2}]
\end{equation*}%
\begin{equation*}
+4(2-T_{n})[\mathrm{Real}(g_{S})\mathrm{Real}(\cos \theta _{L})-\mathrm{Imag}%
(f_{S})\mathrm{Imag}(\sin \theta _{L})]
\end{equation*}%
\begin{equation}
+4T_{n}[\mathrm{Real}(\cos \theta _{L}\sin \theta _{L}^{\ast })\mathrm{Real}%
(f_{S}g_{S}^{\ast })]  \label{tcond}
\end{equation}%
for $i=3,5$ where
\begin{equation}
g_{S}=\left\{
\begin{array}{cc}
(g_{+}+g_{-})/(1+g_{+}g_{-}+f_{+}f_{-}) & \mathrm{TS} \\
(g_{+}+g_{-})/(1+g_{+}g_{-}+f_{+}f_{-}) & \mathrm{USS}%
\end{array}%
\right.  \label{gs}
\end{equation}

\begin{equation}
f_{S}= \left\{
\begin{array}{cc}
i(f_{+}g_{-}-f_{-}g_{+})/(1+g_{+}g_{-}+f_{+}f_{-}) & \mathrm{TS} \\
(f_{+}+f_{-})/(1+g_{+}g_{-}+f_{+}f_{-}) & \mathrm{USS}%
\end{array}
\right.  \label{fs}
\end{equation}

\begin{equation}
\bar{f}_{S}= \left\{
\begin{array}{cc}
(f_{+}+f_{-})/(1+g_{+}g_{-}+f_{+}f_{-}) & \mathrm{TS} \\
i(f_{+}g_{-}- g_{+}f_{-})/(1+g_{+}g_{-}+f_{+}f_{-}) & \mathrm{USS.}%
\end{array}
\right.  \label{bfs}
\end{equation}
Negative sign appears in Eq. (\ref{tcond}), in contrast to Eq. (\ref{cond1}),
since Cooper pairs cannot carry heat.


In the following section, we will discuss the normalized conductance $\sigma
_{T}(T)=\sigma _{S}(T)/\sigma _{N}$, the normalized thermal conductance $%
\kappa _{T}(T)=\kappa (T)/\kappa _{N}(T)$ and the normalized Lorentz ratio $%
L_{T}=\kappa _{T}(T)/\sigma _{T}(T)$ where $\sigma _{N}$ and $\kappa _{N}$
refer to the normal state and are given by $\sigma _{N}=1/(R_{d}+R_{b})$ and
$\kappa _{N}(T)=\pi ^{2}T/(3\left( {R_{b}+R_{d}}\right) e^{2})$ respectively.


\section{Results}

\subsection{$s$-wave case}

\begin{figure}[htb]
\begin{center}
\scalebox{0.4}{
\includegraphics[width=22.0cm,clip]{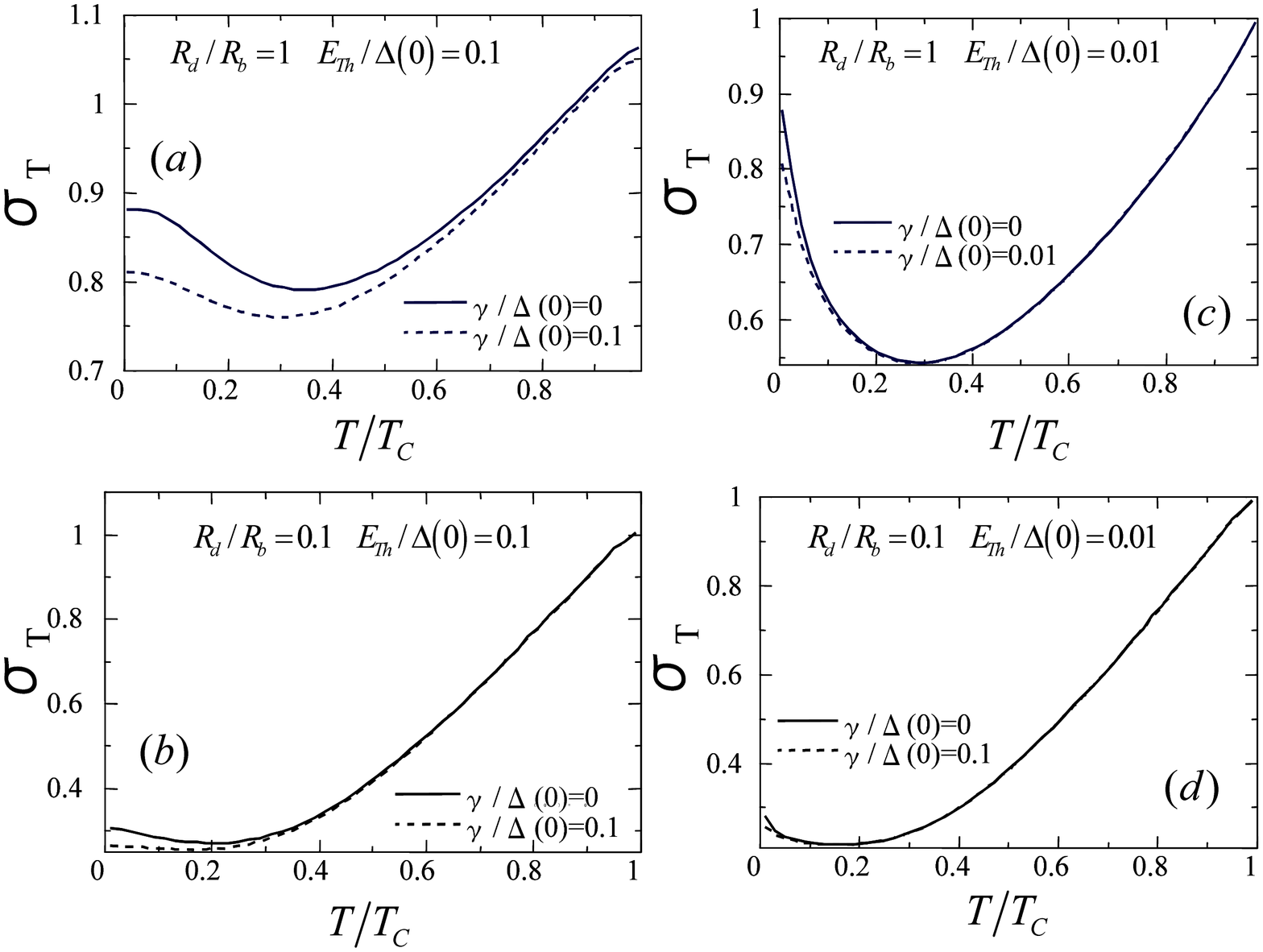}
}
\end{center}
\par
\caption{ Normalized tunneling conductance for $s$-wave superconductor with $
Z=3$. (a) $R_{d}/R_{b}=1$ and $E_{Th}/\Delta
(0)=0.1$. (b) $R_{d}/R_{b}=0.1$ and $E_{Th}/\Delta
(0)=0.1$. (c) $R_{d}/R_{b}=1$ and $E_{Th}/\Delta
(0)=0.01$. (d) $R_{d}/R_{b}=0.1$ and $E_{Th}/\Delta
(0)=0.01$.}
\label{f1}
\end{figure}

First, we study the dependence of electrical conductance at the zero voltage, $%
\sigma _{T}$, on temperature as shown in Fig.~\ref{f1}, where we choose the
relatively strong barrier $Z=3$, $R_{d}/R_{b}=1$ in (a) and (c), and $%
R_{d}/R_{b}=0.1$ in (b) and (d). The Thouless energy is $E_{Th}/\Delta
(0)=0.1$ in (a) and (b), and $E_{Th}/\Delta (0)=0.01$ in (c) and (d).
The tunneling conductance has a peak at $T/T_{C}=0$ and minimum at $%
T/T_{C}\sim 0.3$ as shown in (a)-(d). The coherent AR due to the proximity
effect in DN is responsible for the peak around zero temperature. The
enhancement of the conductance is more pronounced for $R_{d}/R_{b}=1$ in (a)
and (c) than that for $R_{d}/R_{b}=0.1$ in (b) and (d) because the proximity
effect is more prominent for $R_{d}/R_{b}=1$ than $R_{d}/R_{b}=0.1$. The
width of the peak is of the order of $E_{Th}/\Delta (0)$. The magnetic
impurity scattering suppresses the proximity effect. Thus the height of peak
decreases with the increase of $\gamma /\Delta (0)$ as shown in Fig.~\ref{f1}%
. For low transparent interfaces the proximity effect enhances the
conductance around zero temperature. For large $T/T_{C}$, $\sigma _{T}$
increases monotonically with increasing $T/T_{C}$.

\begin{figure}[htb]
\begin{center}
\scalebox{0.4}{
\includegraphics[width=22.0cm,clip]{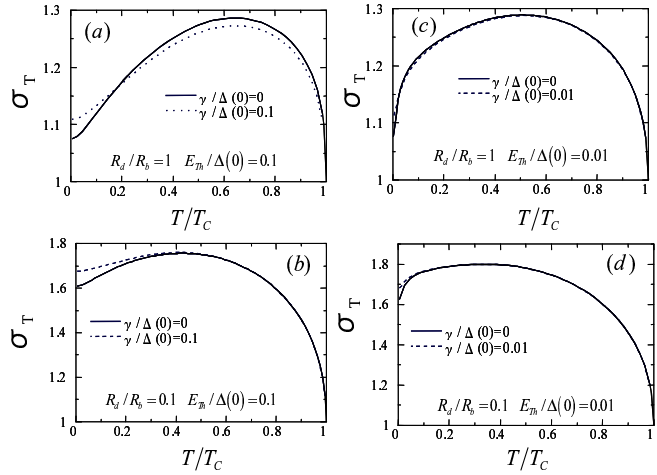}
}
\end{center}
\par
\caption{ Normalized tunneling conductance for $s$-wave superconductor with $%
Z=0$.  (a) $R_{d}/R_{b}=1$ and $E_{Th}/\Delta
(0)=0.1$. (b) $R_{d}/R_{b}=0.1$ and $E_{Th}/\Delta
(0)=0.1$. (c) $R_{d}/R_{b}=1$ and $E_{Th}/\Delta
(0)=0.01$. (d) $R_{d}/R_{b}=0.1$ and $E_{Th}/\Delta
(0)=0.01$.}
\label{f2}
\end{figure}

The corresponding plots for $Z=0$ are shown in Fig.~\ref{f2} (a) - (d) where $
\sigma _{T}$ has a dip at $T/T_{C}=0$ and a maximum at $T/T_{C}\sim 0.5$. It
is known that similar dip-like structures appear also in the conductance as
a function of a bias voltage~\cite{TGK}.
The dip becomes broader for larger magnitudes of $E_{Th}/\Delta (0)$. For
high transparent interfaces, the proximity effect suppresses the conductance
around zero temperature. As a result, magnetic impurity
scattering leads to enhancement of $\sigma _{T}$, as illustrated by our
numerical calculations. The non-monotonic temperature dependence of $\sigma
_{T}$ is a unique feature of diffusive junctions. The structures in Fig.~\ref{f2} are essentially different from those by the VZK theory\cite{Volkov}, which stems from the high transparency at the interface.

\begin{figure}[htb]
\begin{center}
\scalebox{0.4}{
\includegraphics[width=13.0cm,clip]{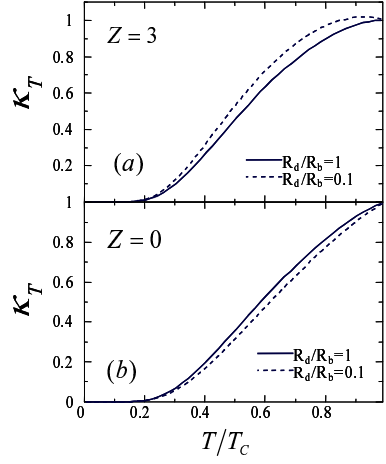}
}
\end{center}
\par
\caption{ Normalized thermal conductance for $s$-wave superconductor with $%
E_{Th}/\Delta(0)=0.1$ and $\protect\gamma/\Delta(0)=0$. (a) $Z=3$ and (b) $Z=0$.}
\label{f3}
\end{figure}

%
Next we study the thermal conductance, $\kappa_{T}$, as a function of
temperature, where $\kappa_{T}$ is normalized by its value
in the normal state, and we fix $E_{Th}/\Delta(0)=0.1$ and $%
\gamma/\Delta(0)=0$. We will show that $\kappa_{T}$ is almost
independent of $E_{Th}/\Delta(0)$ and $\gamma/\Delta(0)$, which
implies that the coherent transmission due to the proximity effect
does not affect the thermal conductance. Both a quasiparicle just on
the Fermi energy and that with finite excitation energy ($\epsilon$)
can carry the electric current. The results in Figs.~\ref{f1} and ~\ref{f2} show
the sensitivity of electrical conductance around the zero
temperature to $E_{Th}/\Delta(0)$ and $\gamma/\Delta(0)$ because the
contribution of a quasiparticle just on the Fermi erengy is governed
by the proximity effect. In the case of thermal conductance, on the
other hand, only quasiparticles with finite energy can carry heat.
At low temperatures, quasiparticles with $\epsilon \sim T \ll T_c$
can contribute to $\kappa_{T}$. Such quasiparticles, however, are
not allowed in the presence of the gap in superconductors. As a
result, $\kappa_{T}$ becomes almost zero around the zero
temperature. In high temperatures such as $T \sim T_C$, only a
quasiparticle with $\epsilon \sim T_c$ contributes to $\kappa_{T}$.
In such energy range, the quasiparticle spectrum in DN is almost
independent of $E_{Th}$ and $\gamma$. Therefore $\kappa_{T}$ is
insensitive to $E_{Th}$ and
$\gamma$. To substantiate it, we show some numerical examples in Fig. \ref%
{f19} where the independence is confirmed. As shown in Fig.~\ref{f3}, $\kappa_T$ increases with increasing $%
T/T_C $ for both $Z=3$ and $Z=0$. Contrary to the case of $\sigma_{T}$, the
magnitude of $\kappa_{T}$ is reduced (enhanced) with the increase of $%
R_{d}/R_{b}$ for $Z=3$ ($Z=0$).


\begin{figure}[htb]
\begin{center}
\scalebox{0.4}{
\includegraphics[width=13.0cm,clip]{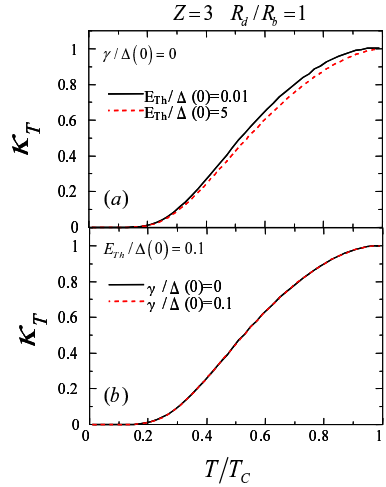}}
\end{center}
\par
\caption{(Color online) Normalized thermal conductance for $s$-wave
superconductor with $Z=3$ and $R_{d}/R_{b}=1$. (a) $\protect\gamma/\Delta(0)=0$ and  (b) $E_{Th}/\Delta
(0)=0.1$.}
\label{f19}
\end{figure}

\begin{figure}[htb]
\begin{center}
\scalebox{0.4}{
\includegraphics[width=13.0cm,clip]{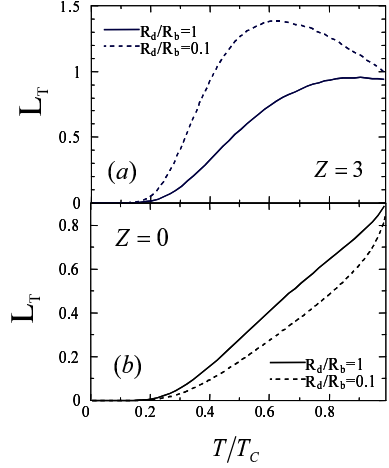}
}
\end{center}
\par
\caption{ Normalized Lorentz ratio for $s$-wave superconductor with $%
E_{Th}/\Delta(0)=0.1$ and $\protect\gamma/\Delta(0)=0$. (a) $Z=3$ and (b) $Z=0$.}
\label{f4}
\end{figure}

We plot the Lorentz ratio, $L_{T}$, in Fig. \ref{f4} for several $R_d/R_b$,
where $E_{Th}/\Delta(0)=0.1$ and $\gamma/\Delta(0)=0$. We confirmed that the
Lorentz ratio is almost independent of $E_{Th}/\Delta (0)$ and $\gamma
/\Delta (0)$. The Lorentz ratio is zero for small $T/T_{C}$ and is linear in
$T/T_{C}$ in the intermediate region. For $Z=3$ and $R_{d}/R_{b}=0.1$, $%
L_{T} $ has a peak at $T/T_{C}\sim 0.7$, whereas it is a monotonic
increasing function of $T/T_{C}$ for $Z=3$ and $R_{d}/R_{b}=1$ (Fig. \ref{f4}%
(a)). For $Z=0$ and $R_{d}/R_{b}=1$, $L_{T}$ is linear in $T/T_{C}$ for $%
T/T_{C}\geq 0.3 $, while  for $Z=0$ and $R_{d}/R_{b}=0.1$ , $%
L_{T}$ is linear in $T/T_{c}$ in the intermediate region (Fig. \ref{f4}(b)).

\begin{figure}[htb]
\begin{center}
\scalebox{0.4}{
\includegraphics[width=13.0cm,clip]{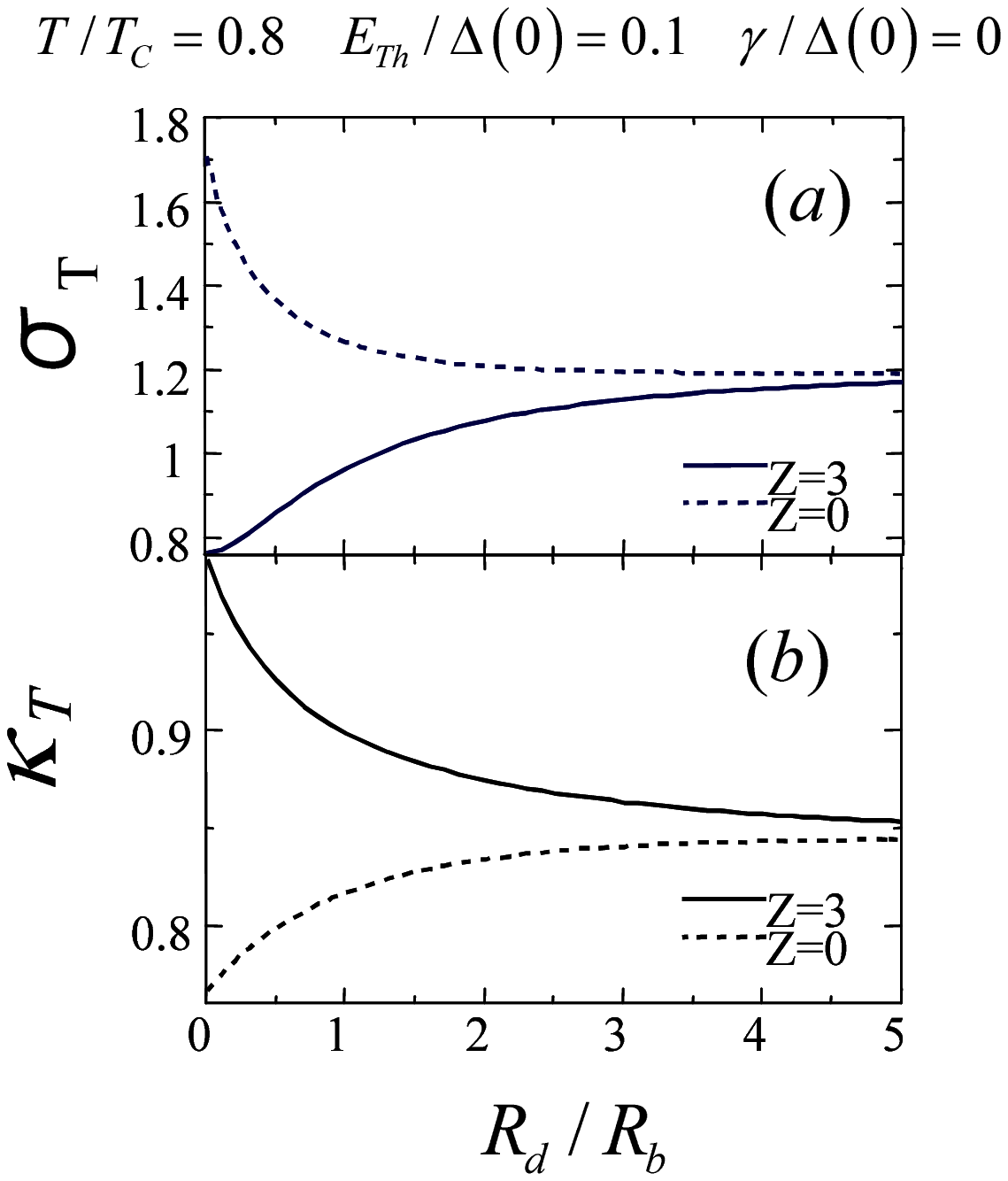}
}
\end{center}
\par
\caption{Normalized tunneling conductance (a) and thermal conductance (b) for $s$%
-wave superconductor with $T/T_C=0.8$, $%
E_{Th}/\Delta(0)=0.1$ and $\protect\gamma/\Delta(0)=0$.}
\label{f5}
\end{figure}

Figure~\ref{f5} shows $R_{d}/R_{b}$ dependence of tunneling conductance and
thermal conductance normalized by their normal values at $T/T_{C}=0.8$,
where $E_{Th}/\Delta (0)=0.1$ and $\gamma /\Delta (0)=0$. In low transparent
interface (i.e., $Z=3$), the probability of AR increases with increasing $%
R_{d}/R_{b}$\cite{Asano}. Thus the magnitude of $\sigma _{T}$ increases with increasing $%
R_{d}/R_{b}$. On the other hand, $\kappa _{T}$ is a decreasing function of $%
R_{d}/R_{b}$. In high transparent interface (i.e., $Z=0$), the proximity
effect suppresses the electrical conductance because the DN plays a role of
the insulating barrier. In this case the probability of AR decreases with
increasing $R_{d}/R_{b}$. Thus $\sigma _{T}$ is a decreasing function of $%
R_{d}/R_{b}$. The results also show that $\kappa _{T}$ is an increasing
function of $R_{d}/R_{b}$. In both (a) and (b), $\sigma _{T}$ and $\kappa
_{T}$ are close to constants independent of $Z$ for sufficiently large $%
R_{d}/R_{b}$.


\subsection{$d$-wave case}

In this subsection, we fix $E_{Th}/\Delta(0)=0.1$ and $\gamma=0$ because $%
\kappa_T$ and $L_T$ are insensitive to these parameters. The pair potentials
$\Delta_{\pm}(T)$ are given by $\Delta_{\pm}(T)=\Delta(T)\cos[2(\phi \mp
\alpha)]$, where $\alpha$ denotes an angle between the normal to the
interface and the crystal axis of $d$-wave superconductors, and $\phi$
denotes an injection angle of a quasiparticle measured from the $x$-axis.
The amplitude of pair potentials, $\Delta(T)$, is the same as that of $s$%
-wave superconductors. We choose $0 \leq \alpha \leq \pi/4$. It is known
that quasiparticles with $\pi/4 - \alpha < \phi < \pi/4 + \alpha $ can
contribute to the MARS at the interface and are responsible for ZBCP in low
transparent junctions. It was shown that the proximity effect and MARS do
not coexist in the $d$-wave symmetry. In fact, at $\alpha=0$, the MARS does
not exist while the proximity effect is possible. On the other hand at $%
\alpha/\pi=0.25$, the proximity effect is not possible, whereas the MARS
appears (see Fig.~\ref{f0} and Ref.~\cite{TNGK}). Thus we can expect similar
results to the $s$-wave symmetry in the case of $\alpha=0$.

\begin{figure}[htb]
\begin{center}
\scalebox{0.4}{
\includegraphics[width=13.0cm,clip]{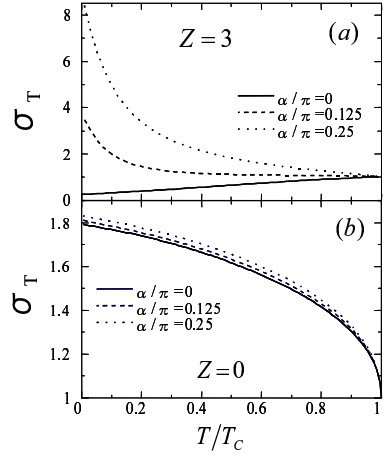}
}
\end{center}
\par
\caption{ Normalized tunneling conductance for $d$-wave superconductor with $R_{d}/R_{b}=0.1$, $%
E_{Th}/\Delta(0)=0.1$ and $\protect\gamma/\Delta(0)=0$. (a) $Z=3$ and (b) $Z=0$.}
\label{f6}
\end{figure}

In Fig.~\ref{f6}, we plot the tunneling conductance as a function of
temperatures for several choices of $\alpha $, where $R_{d}/R_{b}=0.1$. At $Z=3$ in (a), $\sigma _{T}$ for $\alpha /\pi
=0.25 $ and 0.125 increase drastically with decreasing temperatures because
of the resonant transmission through the MARS. Thus such behavior is not
found in $\sigma _{T}$ with $\alpha =0$. In high temperatures, $\sigma _{T}$
for $\alpha /\pi =0$, 0.125 and 0.25 get close together. At $Z=0$, $\sigma
_{T}$ is a monotonic decreasing function of $T/T_{C}$ for all $\alpha $. The
results show that $\sigma _{T}$ slightly increases with $\alpha $ as shown
in (b). The dependence on $\alpha $ of $\sigma _{T}$ at $Z=0$ is very small
in comparison with that at $Z=3$.

\begin{figure}[htb]
\begin{center}
\scalebox{0.4}{
\includegraphics[width=13.0cm,clip]{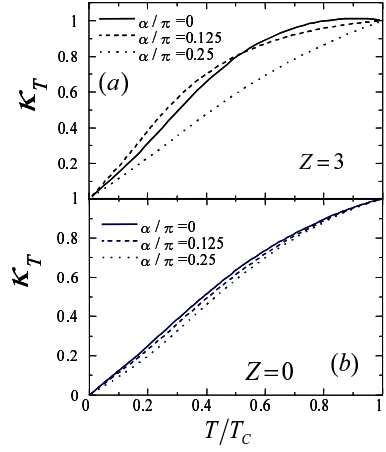}
}
\end{center}
\par
\caption{ Normalized thermal conductance for $d$-wave superconductor with $R_{d}/R_{b}=0.1$, $%
E_{Th}/\Delta(0)=0.1$ and $\protect\gamma/\Delta(0)=0$. (a) $Z=3$ and (b) $Z=0$.}
\label{f7}
\end{figure}

In Fig.~\ref{f7}, we show thermal conductance as a function of temperatures
for several $\alpha$ with $R_d/R_b=0.1$. In
both $Z=3$ and 0, $\kappa_T$ are monotonic increasing function of $T/T_C$
and are proportional to $T/T_C$ for small $T/T_C$. This linear dependence of
$\kappa_T$ for low temperatures is not seen in the $s$-wave symmetry (Fig.~%
\ref{f3}) and reflects the line nodes of the pair potential. The results
also show that $\kappa_T$ depends on $\alpha$ for $Z=3$ in (a), whereas it
is almost independent of $\alpha$ for $Z=0$ in (b).

\begin{figure}[htb]
\begin{center}
\scalebox{0.4}{
\includegraphics[width=13.0cm,clip]{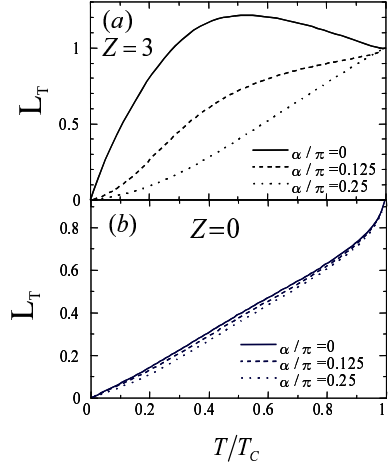}
}
\end{center}
\par
\caption{ Normalized Lorenz ratio for $d$-wave superconductor with $R_{d}/R_{b}=0.1$, $%
E_{Th}/\Delta(0)=0.1$ and $\protect\gamma/\Delta(0)=0$. (a) $Z=3$ and (b) $Z=0$.}
\label{f8}
\end{figure}

In Fig.~\ref{f8}, the Lorentz ratio is shown for several $\alpha$, where $%
R_{d}/R_{b}=0.1$. The Lorentz ratio has a peak
at $T/T_{C}\sim 0.4$ for  $Z=3$ and $\alpha =0$ as shown in (a). This peak
gradually disappears with increasing $\alpha $. For $Z=0$, $L_{T}$ is
proportional to $T/T_{C}$ almost independent of $\alpha$ as shown in (b).

\begin{figure}[htb]
\begin{center}
\scalebox{0.4}{
\includegraphics[width=13.0cm,clip]{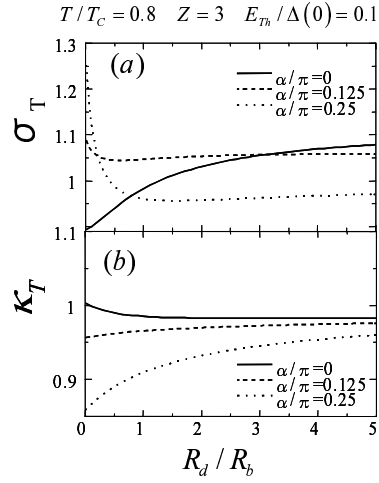}
}
\end{center}
\caption{ Normalized tunneling conductance (a) and thermal conductance (b) for $d$%
-wave superconductor with $T/T_C=0.8$, $Z=3$, $%
E_{Th}/\Delta(0)=0.1$ and $\protect\gamma/\Delta(0)=0$. }
\label{f9}
\end{figure}

\begin{figure}[htb]
\begin{center}
\scalebox{0.4}{
\includegraphics[width=13.0cm,clip]{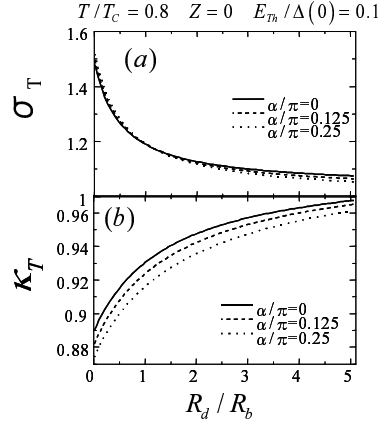}
}
\end{center}
\caption{ Normalized tunneling conductance (a) and thermal conductance (b) for $d$%
-wave superconductor with $T/T_C=0.8$, $Z=0$, $%
E_{Th}/\Delta(0)=0.1$ and $\protect\gamma/\Delta(0)=0$. }
\label{f10}
\end{figure}

Figures~\ref{f9} and \ref{f10} show $R_{d}/R_{b}$  dependence of tunneling conductance and
thermal conductance which are normalized by their normal values at $%
T/T_{C}=0.8$. For $Z=3$ $\sigma _{T}$ is an
increasing function of $R_{d}/R_{b}$ for $\alpha /\pi =0$ as shown in Fig. \ref{f9} (a) because the 
probability of the AR increases with increasing $R_{d}/R_{b}$ as in the $s$%
-wave symmetry. The peaks at $R_{d}/R_{b}=0$ in $\alpha
/\pi =0.25$ and 0.125 is a consequence of the MARS. The impurity scatterings
in DN simply suppress the magnitude of $\sigma _{T}$ at $\alpha /\pi =0.25$
because the the proximity effect is absent in this case. As a result, $%
\sigma _{T}$ becomes a decreasing function of $R_{d}/R_{b}$ at $\alpha /\pi
=0.25$ as shown in Fig. \ref{f9} (a). For sufficiently large $R_{d}/R_{b}$, $%
\sigma _{T}$ is close to a constant. For $\alpha /\pi =0$ $\kappa _{T}$ is a decreasing function
of $R_{d}/R_{b}$ while it is an increasing function of $%
R_{d}/R_{b}$ for $\alpha /\pi =0.125$ and $\alpha /\pi =0.25$ as shown in
Fig. \ref{f9} (b). For sufficiently large $R_{d}/R_{b}$, $\kappa _{T}$ is
close to a constant for all $\alpha $. In Fig.~\ref{f9}(b), we find that the
formation of the MARS suppresses $\kappa _{T}$. This
can be interpreted as follows. When the MARS is formed at the interface, the
quasiparticle density of states becomes large around the zero energy. Such
states around the zero energy, however, can not carry the heat. The zero
energy peak in the density of states means suppression of the density of
states in higher energies which can carry the heat. Thus the formation of
the MARS suppresses the thermal conductance. At $Z=0$ in Fig.~\ref{f10}, the
probability of the AR decreases as increasing $R_{d}/R_{b}$ for all $\alpha $%
. The line shapes of $\sigma _{T}$ and $\kappa _{T}$ are understood in the
same way as those in the $s$-wave case with $Z=0$ in Fig.~\ref{f5}%
.

\subsection{$p$-wave case}

Here we also fix $E_{Th}/\Delta(0)=0.1$ and $\gamma=0$ because $\kappa_T$
and $L_T$ are insensitive to these parameters. We choose the pair potentials
$\Delta_{\pm}$ as $\Delta_{\pm}=\pm\Delta(T)\cos[(\phi \mp \alpha)]$ where $%
\alpha$ denotes an angle between the normal to the interface and the lobe
direction of the $p$-wave pair potential and $\Delta(T)$ is the maximum
amplitude of the pair potential. In the following, we choose $0
\leq \alpha \leq \pi/2$. It is known that quasiparticles with injection
angle $\phi$ with $-\pi/2 + \alpha < \phi < \pi/2 -\alpha $ can contribute
to the formation of the MARS at the interface. In particular at $\alpha=0$,
the MARS and the proximity effect perfectly coexist, which causes the
penetration of the resonant states into the DN~\cite{p-wave}. On the other
hand, for $\alpha/\pi=0.5$, neither the MARS nor the proximity effect exist
(see Fig.~\ref{f0} and Ref.~\onlinecite{p-wave}).

\begin{figure}[htb]
\begin{center}
\scalebox{0.4}{
\includegraphics[width=13.0cm,clip]{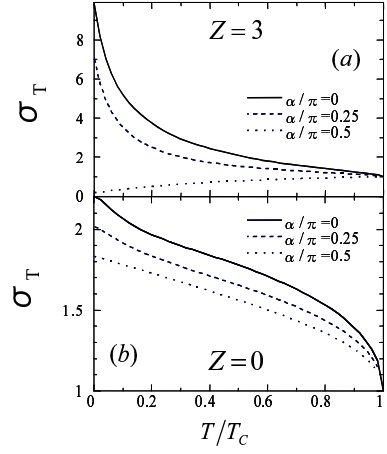}
}
\end{center}
\par
\caption{ Normalized tunneling conductance for $p$-wave superconductor with $R_{d}/R_{b}=0.1$, $%
E_{Th}/\Delta(0)=0.1$ and $\protect\gamma/\Delta(0)=0$. (a) $Z=3$ and (b) $Z=0$.}
\label{f11}
\end{figure}

In Fig. \ref{f11}, we show the calculated results of tunneling conductance
for several $\alpha$, where $R_d/R_b=0.1$. At $%
Z=3 $, $\sigma_T$ for $\alpha/\pi=0$ and 0.25 increases drastically with the
decrease of temperatures because of the resonant transmission via the MARS, while the result for $\alpha/\pi=0.5$ monotonically increases
with $T$. In the case of $Z=0$, $\sigma_T$ increases with decreasing $T$
irrespective of $\alpha$.

\begin{figure}[htb]
\begin{center}
\scalebox{0.4}{
\includegraphics[width=13.0cm,clip]{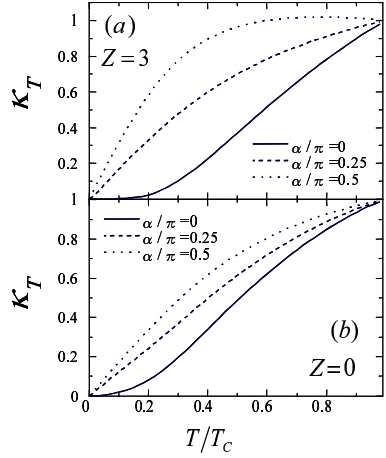}
}
\end{center}
\par
\caption{ Normalized thermal conductance for $p$-wave superconductor with $R_{d}/R_{b}=0.1$, $%
E_{Th}/\Delta(0)=0.1$ and $\protect\gamma/\Delta(0)=0$.  (a) $Z=3$ and (b) $Z=0$.}
\label{f12}
\end{figure}

The thermal conductance in Fig. \ref{f12} is a monotonic increasing function
of $T/T_{C}$, where $R_{d}/R_{b}=0.1$. Except
for $\alpha =0$, $\kappa _{T}$ is proportional to $T$ for small $T/T_{C}$.
This behavior is also found in the $d$-wave symmetry in Fig. \ref{f7} and
stems from the line nodes in the pair potentials. At $\alpha =0$, $\kappa
_{T}$ is expected to be an exponential function of $T/T_{C}$. Thus $\kappa
_{T}$ increases with increasing $\alpha $ as shown in both $Z=3$ and 0.
Although line node of the pair potential exists in this case, $\kappa _{T}$
has an exponential dependence on $T$  as in the $s$-wave case
because the direction of the line node is perpendicular to that of the
thermal current.

\begin{figure}[htb]
\begin{center}
\scalebox{0.4}{
\includegraphics[width=13.0cm,clip]{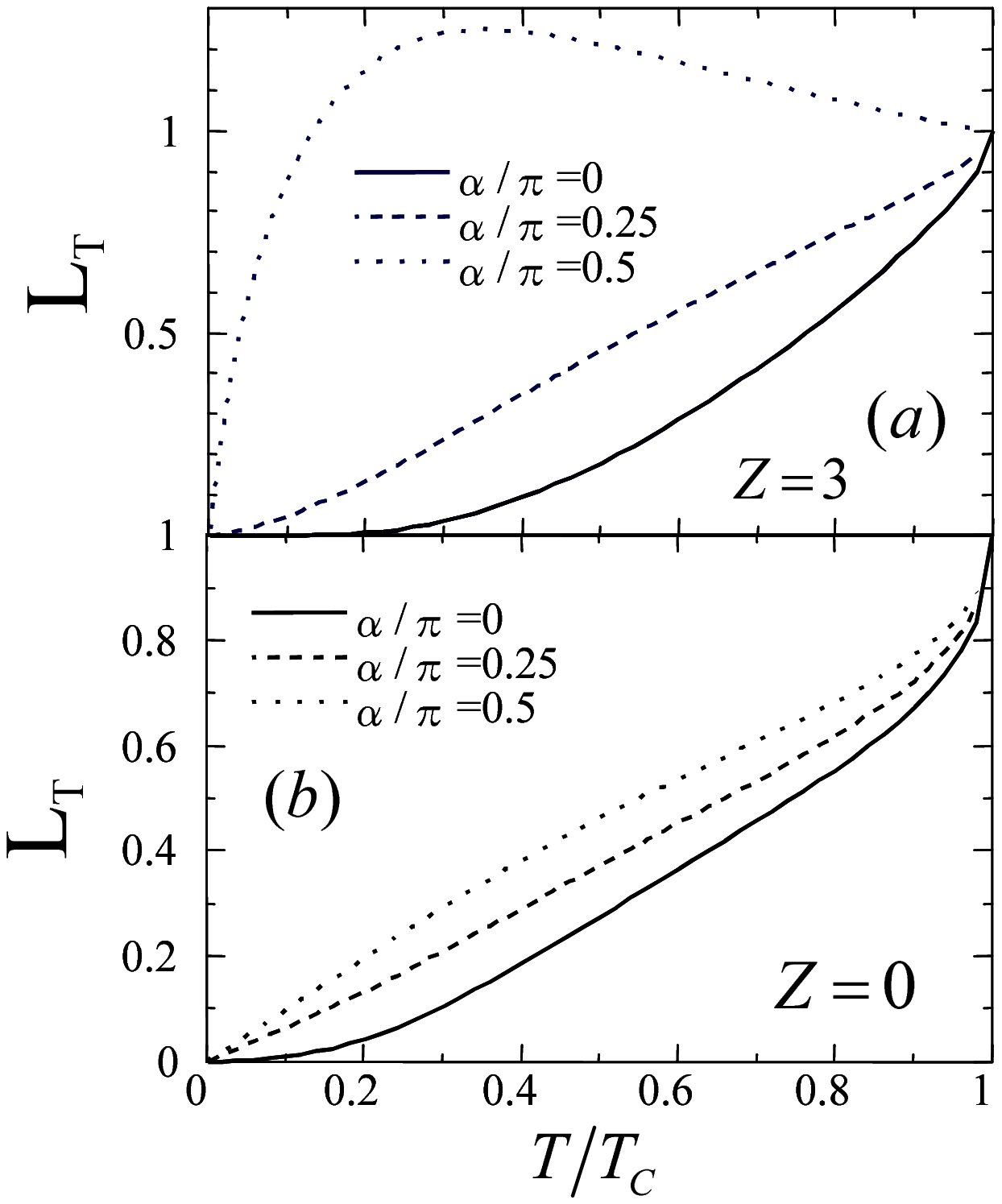}
}
\end{center}
\par
\caption{ Normalized Lorenz ratio for $p$-wave superconductor with $R_{d}/R_{b}=0.1$, $%
E_{Th}/\Delta(0)=0.1$ and $\protect\gamma/\Delta(0)=0$. (a) $Z=3$ and (b) $Z=0$.}
\label{f13}
\end{figure}

The Lorentz ratio has a peak at $T/T_{C}\sim 0.3$ for $\alpha =0$ and $Z=3$%
, as shown in Fig.~\ref{f13}(a). This peak tends to disappear for larger $%
\alpha $. At $Z=0$ $L_{T}$ linearly increases with the increase of temperatures in intermediate temperature regime as shown in (b).

\begin{figure}[htb]
\begin{center}
\scalebox{0.4}{
\includegraphics[width=13.0cm,clip]{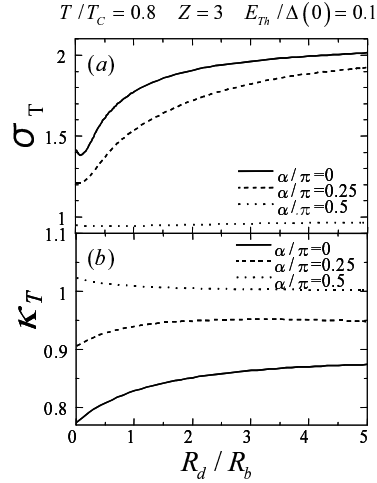}
}
\end{center}
\caption{ Normalized tunneling conductance (a) and thermal conductance (b) for $p$%
-wave superconductor with $T/T_C=0.8$, $Z=3$, $%
E_{Th}/\Delta(0)=0.1$ and $\protect\gamma/\Delta(0)=0$.}
\label{f14}
\end{figure}

\begin{figure}[htb]
\begin{center}
\scalebox{0.4}{
\includegraphics[width=13.0cm,clip]{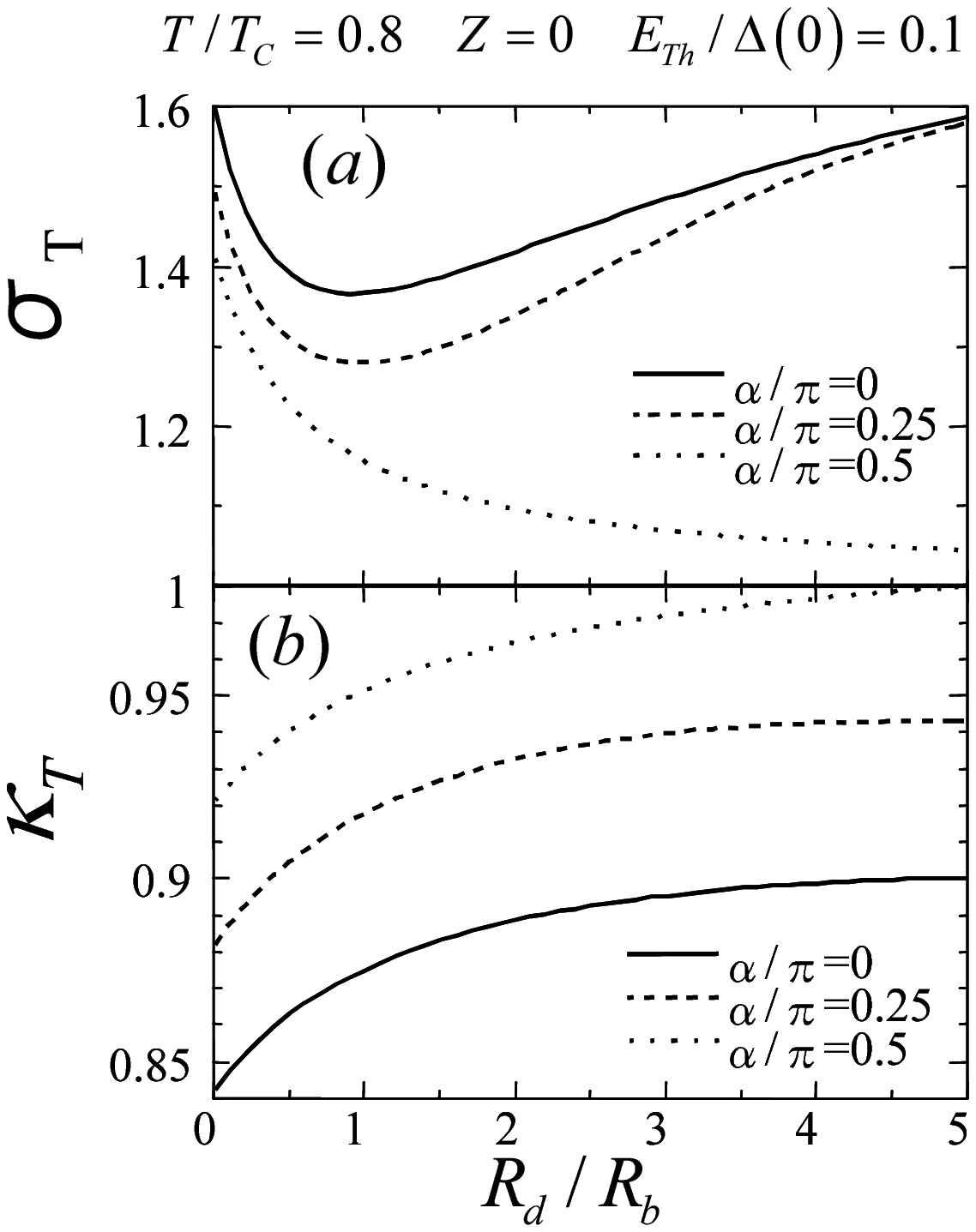}
}
\end{center}
\caption{ Normalized tunneling conductance (a) and thermal conductance (b) for $p$%
-wave superconductor with $T/T_C=0.8$, $Z=0$, $%
E_{Th}/\Delta(0)=0.1$ and $\protect\gamma/\Delta(0)=0$.}
\label{f15}
\end{figure}

Figures~\ref{f14} and \ref{f15} display $R_{d}/R_{b}$ dependence of tunneling
conductance and thermal conductance at $T/T_{C}=0.8$. The results are
normalized by their normal values at $T/T_{C}=0.8$. At $Z=3$, $\sigma _{T}$
has a reentrant behavior in $R_{d}/R_{b}$ for small $\alpha $ as shown in
Fig.~\ref{f14} (a). It is noted that the finite energy states and the zero
energy states contribute to the conductance in qualitatively different ways.
In finite energies, $\sigma _{T}$ decreases  with increasing $%
R_{d}/R_{b} $ because the probability of the AR decreases with increasing $%
R_{d}/R_{b}$. 
On the other hand at the zero energy, $\sigma _{T}$ increase with the
increase of $R_{d}/R_{b}$ due to the formation of the resonant states\cite{p-wave}.
For small (large) $R_{d}/R_{b}$, the contribution of the finite energies
states (the zero energy states) dominates $\sigma _{T}$.
This explains the reentrant behavior in $\sigma _{T}$. In contrast to $%
\alpha =0$, $\sigma _{T}$ for $\alpha =0.5\pi $ is almost constant as shown
in Fig.~\ref{f14} (a) because there are neither the proximity effect nor the
MARS. The characteristics of the thermal conductance can be understood
in a similar way. We note that the zero energy states never contribute to
the thermal transport. At $\alpha =0$, $\kappa _{T}$ increases with
increasing of $R_{d}/R_{b}$ in contrast to the $d$-wave case with $\alpha =0$%
. This difference stems from the existence of the MARS. At $\alpha /\pi =0.5$%
, $\kappa _{T}$ is almost constant  as shown in Fig.~\ref{f14} (b). The
line shapes of $\sigma _{T}$ and $\kappa _{T}$ for $\alpha =0$ and $0.25\pi $
at $Z=0$ in Fig.~\ref{f15} are qualitatively similar to those for $Z=3$. For
$\alpha /\pi =0.5$, $\sigma _{T}$ decreases with $R_{d}/R_{b}$ and $\kappa
_{T}$ increases  with $R_{d}/R_{b}$. These behaviors can be
explained by the fact that the probability of the AR decreases as increasing
$R_{d}/R_{b}$. From Fig.~\ref{f14}(b) for small $R_{d}/R_{b}$, we can also
find that the MARS suppresses $\kappa _{T}$.

\begin{figure}[htb]
\begin{center}
\scalebox{0.4}{
\includegraphics[width=23.0cm,clip]{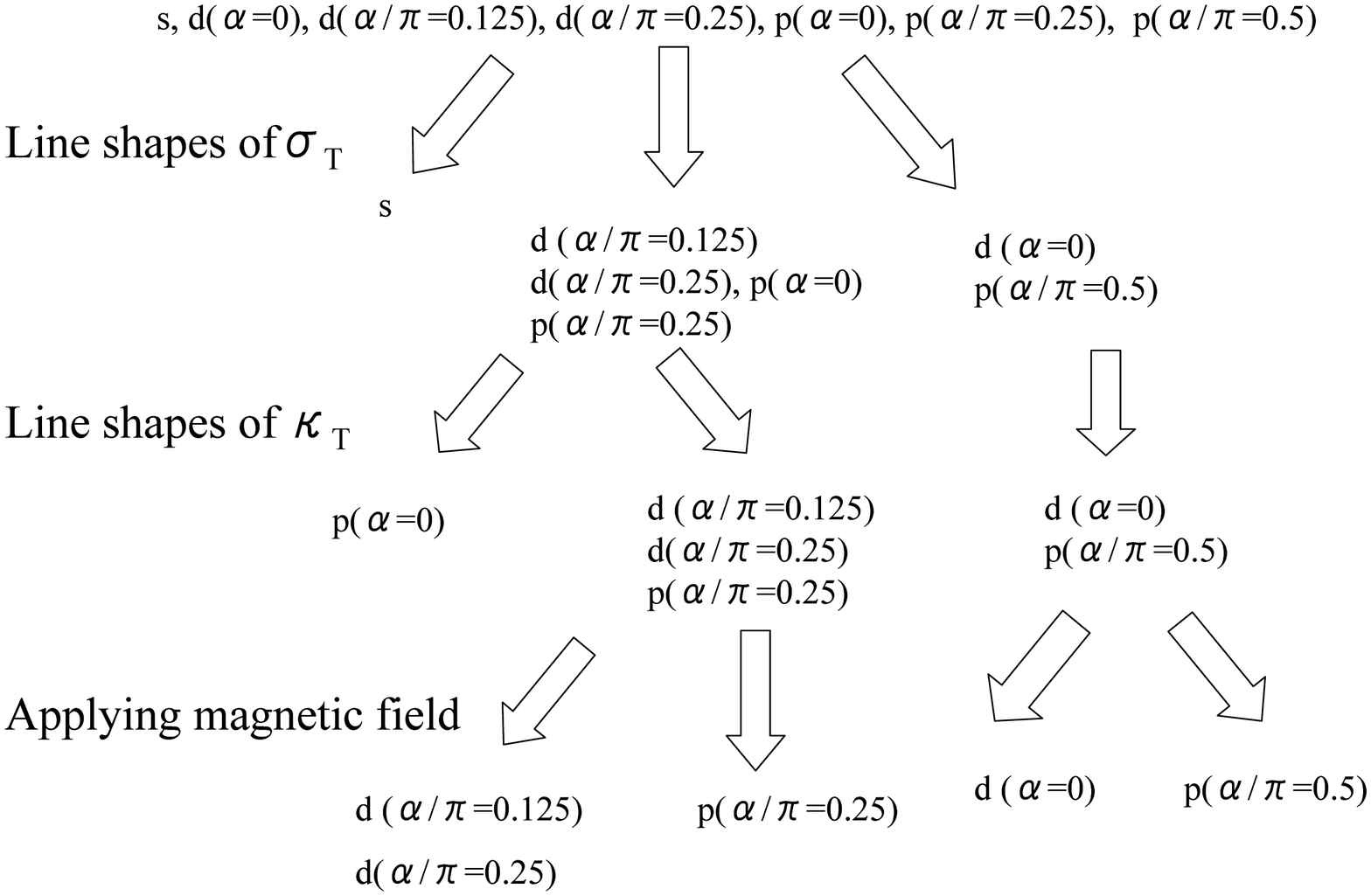}
}
\end{center}
\caption{ Chart for distinguishing $s$-, $d$- and $p$-wave superconductors.}
\label{f18}
\end{figure}

\begin{figure}[htb]
\begin{center}
\scalebox{0.4}{
\includegraphics[width=16.0cm,clip]{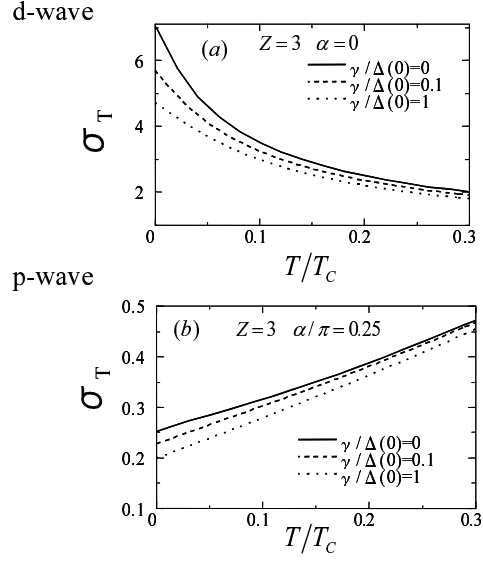}
}
\end{center}
\caption{ Magnetic field dependence of normalized tunneling conductance with $Z=3$ and 
$E_{Th}/\Delta(0)=0.1$ for (a) $d$-wave superconductor with $\alpha /\pi =0$ and (b) $p$-wave superconductor with $\alpha /\pi =0.25$.}
\label{f16}
\end{figure}

\begin{figure}[htb]
\begin{center}
\scalebox{0.4}{
\includegraphics[width=18.0cm,clip]{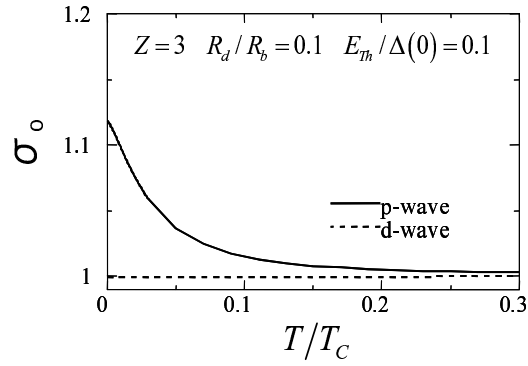}
}
\end{center}
\caption{ The conductance measurable by STS at $x=3L/4$ with $Z=3$, $R_{d}/R_{b}=0.1$, $%
E_{Th}/\Delta(0)=0.1$ and $\protect\gamma/\Delta(0)=0$ for $d$-wave
superconductor with $\protect\alpha/\protect\pi=0.125$ and for $p$-wave
superconductor with $\protect\alpha/\protect\pi=0.25$. }
\label{f17}
\end{figure}

On the basis of calculated results of electrical and thermal conductance, we
propose a way to classify the pairing symmetries with several orientation
angles into six groups as shown in Fig.~\ref{f18}. We have studied seven
junctions: $s$-wave superconductor (s), $d$-wave superconductor with $\alpha
=0$ (d($\alpha =0$)), $d$-wave superconductor with $\alpha /\pi =0.125$ (d($%
\alpha /\pi =0.125$)), $d$-wave superconductor with $\alpha /\pi =0.25$ (d($%
\alpha /\pi =0.25$)), $p$-wave superconductor with $\alpha =0$ (p($\alpha =0$%
)), $p$-wave superconductor with $\alpha /\pi =0.25$ (p($\alpha /\pi =0.25$%
)) and $p$-wave superconductor with $\alpha /\pi =0.5$ (p($\alpha /\pi =0.5$%
)). We focus on the results with $Z=3$ because symmetries of pair potentials
are better characterized by transport properties in lower transparent
junctions. Checking the line shapes of $\sigma _{T}$ and $\kappa _{T}$, we
can separate these junctions into four groups as shown in the first and the
second processes of Fig.~\ref{f18}(see Figs. \ref{f1}, \ref{f3}, \ref{f6}, %
\ref{f7}, \ref{f11}, \ref{f12} and TABLE I). Next applying a weak magnetic
field $H$ parallel to the junction plane, $\sigma _{T}$ for d($\alpha =0$)
and p($\alpha /\pi =0.25$) junctions decreases as increasing magnetic field 
as shown in Fig.~\ref{f16} since the proximity effect is suppressed by the
applied magnetic field~\cite{Belzig1}. On the other hand $\sigma _{T}$ for d(%
$\alpha /\pi =0.125$), d($\alpha /\pi =0.25$) or p($\alpha /\pi =0.5$)
junctions is robust against applied magnetic field since there is almost no
proximity effect (third process in Fig.~\ref{f18}). We note that the
pair-breaking rate $\gamma $ is given by $e^{2}w^{2}DH^{2}/6$, where $w$ is
the transverse size of the DN\cite{Belzig1}. Assuming $w=10^{-5}m$, $%
D=10^{-3}m^{2}/s$, $\Delta (0)=10^{-3}eV$, and $H=10^{-2}T$, we can
estimate the pair-breaking rate $\gamma /\Delta (0)\sim 1$.
Distinguishing p($\alpha /\pi =0.25$) from d($\alpha /\pi =0.125$)
is a delicate problem since both of them have the proximity effect
and the MARS. The two junctions, however,
have qualitative difference in the density of states (DOS) in DN. In p($%
\alpha /\pi =0.25$) junctions, the DOS has a  zero
 energy  peak due to the formation of the resonant
states whereas the DOS in the d($\alpha /\pi =0.125$)
junctions does not show such zero energy peak~\cite{p-wave}. As a result, the conductance measurable by scanning tunneling spectroscopy (STS), $\sigma_0$, reflects these features. Here $\sigma_0$ is defined as 
\begin{equation}
\sigma_0 = \frac{1}{2T}\int_{0}^{\infty }\frac{\text{Re}\cos \theta d\epsilon }{\cosh
^{2}[\frac{\epsilon }{2T}]}
\end{equation}%
 and  $\text{Re}\cos \theta$ is the DOS normalized by its normal states value. 
We plot it at $x=3L/4$  as a function of $T/T_{C}$ in Fig.
\ref{f17}. In p($\alpha /\pi =0.25$) junctions, a peak appears at zero temperature in contrast to the case of the d($\alpha /\pi =0.125$)
junctions. Thus we can easily
distinguish these superconductors by STS. 

\begin{table}[tbp]
\caption{Low temperature dependences of $\protect\sigma_T$ and $\protect%
\kappa_T$.}
\label{tab:table1}%
\begin{tabular}{|c|c|c|}
\hline
pairing symmetry & $\sigma_T$ & $\kappa_T$ \\ \hline
s & reentrant & exponential \\ \hline
d($\alpha=0$) & linear & linear \\ \hline
d($\alpha/\pi=0.125$) & inverse & linear \\ \hline
d($\alpha/\pi=0.25$) & inverse & linear \\ \hline
p($\alpha=0$) & inverse & exponential \\ \hline
p($\alpha/\pi=0.25$) & inverse & linear \\ \hline
p($\alpha/\pi=0.5$) & linear & linear \\ \hline
\end{tabular}%
\end{table}

\clearpage

\section{Conclusions}

In the present paper, we have derived a general expression of the thermal
conductance in normal metal / superconductor junctions based on the Usadel
equation under the generalized boundary condition. We have studied the
electrical and thermal transport in diffusive normal metal / $s$-, $d$- and $%
p$-wave superconductor junctions in the presence of magnetic impurities in
normal metals. The main conclusions are summarized as follows.

1. The proximity effect does not influence the thermal conductance.
This statement is illustrated with some numerically calculated
examples.

2. The midgap Andreev resonant states in $d$- or $p$-wave superconductor
junctions suppress the thermal conductance. The formation of MARS
drastically gathers the density of states at the Fermi energy near the DN/S
interface. Such quasiparticles, however, do not carry heat because
excitation energies of them are almost zero.

3. The thermal conductance of the junctions reflects the existence of the
line nodes of the pair potential except for the case that the direction of
the line node is perpendicular to that of the thermal current.

Electric conductance, thermal conductance, and their Lorentz ratio
calculated as a function of temperature depend strongly on a pairing
symmetry of a superconductor. This fact indicates a possibility of
distinguishing one pairing symmetry from another by careful
comparison of the present calculations and experimental results.

In this paper, we have focused on N/S junctions of unconventional
superconductors. So far an extension of the circuit theory to long diffusive
S/N/S junctions has been performed by Bezuglyi et al.~\cite{Bezuglyi2} in $%
s $-wave symmetry. In S/N/S junctions, the multiple AR produces
subharmonic gap structures in I-V
curves~\cite{M1,M2,M3,M4,M5,M6,M7,M8}. In S/N/S junctions of
unconventional superconductors, it is known that MARS leads to the
anomalous current-phase relation and temperature dependence of the
Josephson current~\cite{TKJ}. Effects of unconventional
superconductivity of such I-V curves are an important future issue.
The research in this direction is now in progress and the results
will be reported elsewhere.

%
The authors appreciate useful and fruitful discussions with J.~Inoue,
Yu.~Nazarov and H.~Itoh. This work was supported by NAREGI Nanoscience
Project, the Ministry of Education, Culture, Sports, Science and Technology,
Japan, the Core Research for Evolutional Science and Technology (CREST) of
the Japan Science and Technology Corporation (JST) and a Grant-in-Aid for
the 21st Century COE "Frontiers of Computational Science" . The
computational aspect of this work has been performed at the Research Center
for Computational Science, Okazaki National Research Institutes and the
facilities of the Supercomputer Center, Institute for Solid State Physics,
University of Tokyo and the Computer Center.

\section{Appendix}

Here we calculate the matrix current for the calculation of the thermal
conductance of $d$-wave junctions. We denote $\check{H}_{+} $, $\check{H}%
_{-} $, $\check{B}_{n}$, $\check{I}$ as follows,
\begin{eqnarray}
\check{H}_{+} =\left(
\begin{array}{cc}
\hat{R}_{p} & \hat{K}_{p} \\
0 & \hat{A}_{p}%
\end{array}%
\right) , \ \ \check{H}_{-} =\left(
\begin{array}{cc}
\hat{R}_{m} & \hat{K}_{m} \\
0 & \hat{A}_{m}%
\end{array}%
\right),
\end{eqnarray}

\begin{equation*}
\check{B}_{n}=\left(
\begin{array}{cc}
\hat{B}_{R} & \hat{B}_{K} \\
0 & \hat{B}_{A}%
\end{array}%
\right) ,\ \ \check{I}=\left(
\begin{array}{cc}
\hat{I}_{R} & \hat{I}_{K} \\
0 & \hat{I}_{A}%
\end{array}%
\right)
\end{equation*}%
with $\check{H}_{\pm}=(\check{G}_{2+} \pm \check{G}_{2-})/2.$ In singlet
superconductors, we can choose $\hat{R}_{1}=\cos \theta (x)\hat{\tau}%
_{3}+\sin \theta (x)\hat{\tau}_{2}$ to satisfy the boundary condition at the
interface.

The matrix current is given by $\check{I}=\mathrm{Tr}_{n}[\check{I_n}]$ with
$\check{I_n}=2[\check{G}_{1},\check{B}_{n}]$ and
\begin{eqnarray}
\check{B}_{n}=(-T_{1n}[\check{G}_{1},\check{H}_{-}^{-1}]+\check{H}_{-}^{-1}%
\check{H}_{+}-T_{1n}^{2}\check{G}_{1}\check{H}_{-}^{-1}\check{H}_{+}\check{G}%
_{1})^{-1}(T_{1n}(1-\check{H}_{-}^{-1})+T_{1n}^{2}\check{G}_{1}\check{H}%
_{-}^{-1}\check{H}_{+})
\end{eqnarray}%
where
\begin{equation*}
T_{1n}=\frac{T_{n}}{2-T_{n} + 2\sqrt{1-T_{n}}}.
\end{equation*}
Next, we focus on the Keldysh component. We define $I_{b}$
\begin{equation}
I_{b}=\frac{1}{2}\sum_{n}\mathrm{Tr}[\hat{I}_{K}].
\end{equation}%
After straightforward calculations, $I_{b}$ is given by
\begin{equation*}
I_{b}=\frac{1}{2}\sum_{n}\mathrm{Tr}[(\hat{R}_{1}-\hat{A}_{1})\hat{B}%
_{K}+f_{l}(L_-)(\hat{R}_{1}-\hat{A}_{1})(\hat{B}_{A}-\hat{B}_{R})].
\end{equation*}%
It is necessary to obtain $\hat{B}_{K}$ which is given by
\begin{equation}
\hat{B}_{K}=\hat{D}_{R}^{-1}\hat{N}_{K}-\hat{D}_{R}^{-1}\hat{D}_{K}\hat{D}%
_{A}^{-1}\hat{N}_{A},
\end{equation}
with
\begin{equation*}
\check{D}= -T_{1n}[\check{G}_{1},\check{H}_{-}^{-1}] + \check{H}_{-}^{-1}%
\check{H}_{+} -T_{1n}^{2}\check{G}_{1}\check{H}_{-}^{-1}\check{H}_{+}\check{G%
}_{1}, \ \ \check{D}=\left(
\begin{array}{cc}
\hat{D}_{R} & \hat{D}_{K} \\
0 & \hat{D}_{A}%
\end{array}%
\right),
\end{equation*}
where $\hat{N}_{K}$ and $\hat{N}_{A}$ is the Keldysh and advanced part of $%
\check{N}$ given by
\begin{equation*}
\check{N}=T_{1n}-T_{1n}\check{H}_{-}^{-1}+T_{1n}^{2}\check{G}_{1}\check{H}%
_{-}^{-1}\check{H}_{+} \ \ \check{N}=\left(
\begin{array}{cc}
\hat{N}_{R} & \hat{N}_{K} \\
0 & \hat{N}_{A}%
\end{array}%
\right).
\end{equation*}%
We can express $\hat{N}_{K}$ and $\hat{D}_{K}$ as linear combination of
distribution functions $f_{S}$, $f_{l}(L_-)$, and $f_{t}(L_-)$ as follows,
\begin{equation*}
\hat{N}_{K}=\hat{C}_{1}f_{S}+\hat{C}_{2}f_{l}(L_-)+\hat{C}_{3}f_{t}(L_-),
\end{equation*}%
\begin{equation*}
\hat{D}_{K}=\hat{C}_{4}f_{S}+\hat{C}_{5}f_{l}(L_-)+\hat{C}_{6}f_{t}(L_-),
\end{equation*}%
by $2\times 2$ matrix $C_{i}$ $(i=1-6)$. Taking account of the fact that $%
\hat{D}_{R}^{-1}\hat{C}_{1}$, $\hat{D}_{R}^{-1}\hat{C}_{2}$, $\hat{D}%
_{R}^{-1}\hat{C}_{4}\hat{D}_{A}^{-1}\hat{N}_{A}$ and $\hat{D}_{R}^{-1}\hat{C}%
_{5}\hat{D}_{A}^{-1}\hat{N}_{A}$ can be expressed by the linear combination
of $\hat{\tau}_{2}$ and $\hat{\tau}_{3}$, while $\hat{D}_{R}^{-1}\hat{C}_{3}$
and $\hat{D}_{R}^{-1}\hat{C}_{6}\hat{D}_{A}^{-1}\hat{N}_{A}$ are
proportional to the linear combination of $\hat{1}$ and $\hat{\tau}_{1}$, we
can express $I_{b}$ as follows,
\begin{equation*}
I_{b}=\frac{1}{2}\sum_{n}\mathrm{Tr}[f_S{(\hat{R}_{1}-\hat{A}_{1})(\hat{D}%
_{R}^{-1}\hat{C}_{1}-\hat{D}_{R}^{-1}\hat{C} _{4}\hat{D}_{A}^{-1}\hat{N}_{A})%
}+f_l(L_-){(\hat{R}_{1}-\hat{A}_{1})(\hat{D}_{R}^{-1}\hat{C}_{2}-\hat{D}%
_{R}^{-1}\hat{C} _{5}\hat{D}_{A}^{-1}\hat{N}_{A})+(\hat{R}_{1}-\hat{A}_{1})(%
\hat{B}_{A}-\hat{B}_{R})}]
\end{equation*}%
with
\begin{equation*}
\hat{C}_{1}=T_{1n}[\hat{A}_{m}^{-1}-\hat{R}_{m}^{-1}+T_{1n}(\hat{R}_{1}\hat{R%
}_{m}^{-1}(\hat{R}_{p}-\hat{A}_{p})-\hat{R}_{1}(\hat{A}_{m}^{-1}-\hat{R}%
_{m}^{-1})\hat{A}_{p})]
\end{equation*}

\begin{equation*}
\hat{C}_{2}=T_{1n}^{2}(\hat{R}_{1}-\hat{A}_{1})\hat{A}_{m}^{-1}\hat{A}_{p}
\end{equation*}

\begin{equation*}
\hat{C}_{4}=T_{1n}[\hat{R}_{1}(\hat{A}_{m}^{-1}-\hat{R}_{m}^{-1})-(\hat{A}%
_{m}^{-1}-\hat{R}_{m}^{-1})\hat{A}_{1}]+\hat{R}_{m}^{-1}\hat{R}_{p}-\hat{A}%
_{m}^{-1}\hat{A}_{p}+T_{1n}^{2}[-\hat{R}_{1}\hat{R}_{m}^{-1}(\hat{R}_{p}-%
\hat{A}_{p})\hat{A}_{1}+\hat{R}_{1}(\hat{A}_{m}^{-1}-\hat{R}_{m}^{-1})\hat{A}%
_{p}\hat{A}_{1}]
\end{equation*}

\begin{equation}
\hat{C}_{5}=T_{1n}[-(\hat{R}_{1}-\hat{A}_{1})\hat{A}_{m}^{-1}+\hat{R}%
_{m}^{-1}(\hat{R}_{1}-\hat{A}_{1})-T_{1n}(\hat{R}_{1}\hat{R}_{m}^{-1}\hat{R}%
_{p}(\hat{R}_{1}-\hat{A}_{1})+(\hat{R}_{1}-\hat{A}_{1})\hat{A}_{m}\hat{A}%
_{p}^{-1}\hat{A}_{1})].
\end{equation}

We use the following equations 
\begin{equation*}
\hat{D}_{R}^{-1}(T_{1n}-T_{1n}\hat{R}_{m}^{-1}+T_{1n}^{2}\hat{R}_{1}\hat{R}%
_{m}^{-1}\hat{R}_{p})=\hat{B}_{R},
\end{equation*}%
\begin{equation}
\hat{B}_{R}(1+\hat{R}_{m}^{-1}+T_{1n}\hat{R}_{1}\hat{R}_{p}\hat{R}%
_{m}^{-1})=T_{1n}\hat{R}_{m}^{-1}\hat{R}_{p}
\end{equation}%
where $\hat{B}_{R}$ is given as $\hat{B}_{R}=b_{1}\hat{\tau}_{1}+b_{2}\hat{%
\tau}_{2}+b_{3}\hat{\tau}_{3}$ with
\begin{equation*}
b_{1}=\frac{-iT_{1n}(f_{+}g_{-}-f_{-}g_{+})}{
(1+T_{1n}^{2})(1+g_{+}g_{-}+f_{+}f_{-})+2T_{1n}[\cos \theta
_{L}(g_{+}+g_{-})+\sin \theta _{L}(f_{+}+f_{-})]},
\end{equation*}%
\begin{equation*}
b_{2}=\frac{-T_{1n}[T_{1n}\sin \theta
_{L}(1+g_{+}g_{-}+f_{+}f_{-})+f_{+}+f_{-}]}{%
(1+T_{1n}^{2})(1+g_{+}g_{-}+f_{+}f_{-})+2T_{1n}[\cos \theta
_{L}(g_{+}+g_{-})+\sin \theta _{L}(f_{+}+f_{-})]},
\end{equation*}%
\begin{equation}
b_{3}=\frac{-T_{1n}[T_{1n}\cos \theta
_{L}(1+g_{+}g_{-}+f_{+}f_{-})+g_{+}+g_{-}]}{%
(1+T_{1n}^{2})(1+g_{+}g_{-}+f_{+}f_{-})+2T_{1n}[\cos \theta
_{L}(g_{+}+g_{-})+\sin \theta _{L}(f_{+}+f_{-})]}.
\end{equation}

Finally we reach the expression of $I_{b}$ given by the following equation.
\begin{equation}
I_{b}=\sum_{n}\frac{T_{n}}{2}\frac{C_{0}^{\prime }(f_{s}-f_{l}(L_{-}))}{\mid
(2-T_{n})(1+g_{+}g_{-}+f_{+}f_{-})+T_{n}[\cos \theta _{L}(g_{+}+g_{-})+\sin
\theta _{L}(f_{+}+f_{-})]\mid ^{2}}
\end{equation}%
\begin{equation*}
C_{0}^{\prime }=T_{n}(1+\mid \cos \theta _{L}\mid ^{2}-\mid \sin \theta
_{L}\mid ^{2})(\mid g_{+}+g_{-}\mid ^{2}-\mid f_{+}+f_{-}\mid ^{2}+\mid
1+f_{+}f_{-}+g_{+}g_{-}\mid ^{2}-\mid f_{+}g_{-}-g_{+}f_{-}\mid ^{2})
\end{equation*}%
\begin{equation*}
+4(2-T_{n})[\mathrm{Real}[{(1+g_{+}^{\ast }g_{-}^{\ast }+f_{+}^{\ast
}f_{-}^{\ast })(g_{+}+g_{-})}]\mathrm{Real}(\cos \theta _{L})-\mathrm{Imag}%
(\sin \theta _{L})\mathrm{Imag}[{(1+g_{+}^{\ast }g_{-}^{\ast }+f_{+}^{\ast
}f_{-}^{\ast })(f_{+}+f_{-})}]]
\end{equation*}%
\begin{equation*}
+4T_{n}\mathrm{Real}(\cos \theta _{L}\sin \theta _{L}^{\ast })\mathrm{Real}%
[(f_{+}+f_{-})(g_{+}^{\ast }+g_{-}^{\ast })].
\end{equation*}%
This is a general expression applicable to any singlet superconductors
without broken time reversal symmetry state.

%


\end{document}